\documentclass[10pt]{article}
\usepackage{amssymb,amsmath} 
\usepackage{graphicx}
\usepackage[colorlinks=true, pdfstartview=FitV, linkcolor=blue, citecolor=blue, urlcolor=blue]{hyperref}
\usepackage{hyperref}
\usepackage{bbm}

\def\div{\nabla\cdot } 
\def\pl{\partial}

\def\={\equiv}

\newcommand{\xt}{(\3x,t)}

\newcommand{\ox}{(\3x)}

\newcommand{\qed}{\blacksquare}
\newtheorem{defin}{Definition}
\newtheorem{prop}{Proposition}
\newtheorem{thm}{Theorem}

\newcommand{\bib}{\bibitem}

\newcommand{\nt}{\notag}
\newcommand{\ci}{\cite}
\newcommand{\lab}{\label}

\newcommand{\eq}{\eqref}

\newcommand{\bx}[1]{\boxed{\ #1\ }}

\newcommand{\lp}{\left(}
\newcommand{\rp}{ \right)}
\newcommand{\lb}{ \left[}
\newcommand{\rb}{\right]}

\newcommand{\LB}{\left\lbrace}
\newcommand{\RB}{\right\rbrace}

\newcommand{\harr}[1]{\smash{\mathop{\hbox to .5in{\ \rightarrowfill\ }}
      \limits^{#1}}}

\newcommand{\0}[1]{{(#1)}}

\newcommand{\2}[1]{{\tilde #1}}
\newcommand{\3}[1]{{\boldsymbol #1}}

\newcommand{\bh}[1]{\mbox{\boldmath{{$\rm\hat#1$}}}}
\newcommand{\bt}[1]{{\bf{\tilde #1}}}

\def\a{\alpha} 
\def\b{\beta}

\def\d{\delta} 
\def\e{\varepsilon} 

\def\f{\phi} 
\def\vf{\varphi} 
\def\g{\gamma}
\def\h{\eta} 

\def\k{\kappa}

\def\l{\lambda} 
\def\m{\mu} 
\def\n{\nu}
\def\o{\omega} 
\def\p{\pi} 

\def\q{\theta} 
\def\vq{\vartheta} 
\def\r{\rho}
\def\vr{\varrho} 
\def\s{{\sigma}} 

\def\t{\tau} 
 
\def\x{\xi}
\def\y{\psi} 
\def\z{\zeta}

\def\D{\Delta}

\def\O{\Omega}

\def\Y{\Psi}

\newcommand{\bul}{$\bullet\ $}

\newcommand{\db}{{\,{\rm d}\kern-1.6ex-}}
\newcommand{\dir}{{\pl\kern-1.2ex {/}}}
\newcommand{\dd}{{\rm d}}

\newcommand{\app}{\approx} 
\newcommand{\cc}[1]{{{\mathbb C\hskip.5pt}^{#1}}}

\newcommand{\curl}{\nabla\times}

\newcommand{\grad}{\nabla}

\newcommand{\ie}{{\it i.e., }}

\def\iff{\ \Leftrightarrow\ }
\newcommand{\im}{{\,\rm Im}\ }  
\newcommand{\imp}{\ \Rightarrow\ }
\newcommand{\inv}{^{-1}}

\newcommand{\lra}{\leftrightarrow}

\newcommand{\plra}{\pl^{\kern-1.25ex^\lra}}
\newcommand{\qq}{\quad} 
\newcommand{\qqq}{\qquad} 
\newcommand{\re}{{\,\rm Re}\  }   

\newcommand{\rr}[1]{{{\mathbb R}^{#1}}}

\newcommand{\sgn}{{\,\rm Sgn \,}}
\newcommand{\sh}[1]{\hskip#1ex} 
\newcommand{\sr}{\sqrt}

\newcommand{\sv}[1]{\vskip#1ex}

\def\XXint#1#2#3{{\setbox0=\hbox{$#1{#2#3}{\int}$}
     \vcenter{\hbox{$#2#3$}}\kern-.5\wd0}}

\def\bib#1{\bibitem[#1]{#1}}

\begin{document}

\title{Coherent electromagnetic wavelets and their twisting null congruences}

\author{Gerald Kaiser\\
Center for Signals and Waves\\ Austin, TX\\
kaiser@wavelets.com  $\bullet$\ www.wavelets.com}

\maketitle

\begin{abstract}\noindent 
We construct an electromagnetic field whose scalar potential is a \sl pulsed-beam wavelet \rm $\2\Y$ (an analytic continuation of a classical Huygens wave\-let). The vector potential $\bt A$ is determined up to three complex parameters by requiring that (a) it satisfies the Lorenz gauge condition with $\2\Y$, (b) its current density is supported on the same disk $\5D$ as the charge density of $\2\Y$, (c) it is axisymmetric, and (d) it has the same retarded time dependence as $\2\Y$. By choosing one of the parameters in $\bt A$ appropriately, the electromagnetic field generated by the four-potential $(\bt A, \2\Y)$ can be made \sl null, \rm meaning that $\3E^2=\3B^2$ and $\3E\cdot\3B=0$. We call such fields \sl coherent \rm because upon being radiated, they do not loiter around the source, generating \sl electromagnetic inertia, \rm but immediately propagate out at the speed of light. The coherent EM wavelets define a \sl twisting null congruence of light rays \rm  in Minkowski space, which we show to be identical to the Kerr congruence associated with the Kerr-Newman metric. The latter represents a black hole due to a time-independent charge-current density on a massive disk $\5D$ spinning at the angular velocity $c/a$, where $a$ is the radius of $\5D$. By contrast, our coherent wavelets are electromagnetic pulsed beams radiated by  \sl pulsed \rm charge-current distributions on $\5D$, still spinning at the uniform rate $c/a$.  
\end{abstract}

\tableofcontents

\section{The inertia density of an electromagnetic field}\label{S:inertia}

The energy density $u$ and the Poynting vector $\3S$ of an electromagnetic field $(\3E,\3B)$ in vacuum are given by
\begin{align}\lab{ES}
u\0x=\frac12(\3E\0x^2+\3B\0x^2)\ \ \hbox{and}\ \ \3S\0x=\3E\0x\times\3B\0x,
\end{align}
where $x=\xt\in\rr4$ are the spacetime variables. We are using Heaviside-Lorentz units $(\e_0=\m_0=1)$. $\3S$ satisfies the vector identity
\begin{align*}
\3S^2=\3E^2\3B^2-(\3E\cdot\3B)^2,
\end{align*}
which implies that
\begin{align}\lab{pos}
u^2-\3S^2=\frac14(\3E^2-\3B^2)^2+(\3E\cdot\3B)^2\ge 0.
\end{align}
The \sl electromagnetic momentum density \rm is given by\footnote{In SI units \ci[page 261]{J99}, $\3g=\3S/c^2$.
}
\begin{align*}
\3g\0x=\3S\0x/c\,,
\end{align*}
where $c$ is the speed of light.
For a relativistic \sl particle \rm with energy $E$ and momentum $\3p$, the mass and velocity are given by
\begin{align}\lab{Epv}
m^2c^4=E^2-c^2\3p^2\ge 0\ \ \hbox{and}\ \ \3v=\frac{c^2\3p}E\imp \3v^2\le c^2.
\end{align}
Thus it makes sense to define the \sl electromagnetic inertia density \rm 
\begin{align}\lab{I}
\bx{\5I\0x=c^{-2}\sr{u\0x^2-c^2\3g\0x^2}=\frac1{2c^2}\sr{(\3E^2-\3B^2)^2+4(\3E\cdot\3B)^2}}
\end{align}
and the \sl electromagnetic energy flow velocity \rm 
\begin{align}\lab{vel}
\bx{\3v\0x=\frac{c\3S\0x}{u\0x},\qq  \3v\0x^2\le c^2.}
\end{align}
$\5I\0x$ and $\3v\0x$ are \sl local spacetime fields.  \rm Since $\3E^2-\3B^2$ and $\3E\cdot\3B$ are Lorentz-invariant, $\5I\0x$ is a \sl scalar \rm  field. 

From here on we set
\begin{align*}
c\=1, 
\end{align*}
although $c$ will be inserted into expressions occasionally to clarify their physical contents.

The energy, momentum and inertia densities of a field can be expressed succinctly in terms of the complex vector fields
\begin{align}\lab{F0}
\3F\0x=\3E\0x+i\3B\0x
\end{align}
as follows:
\begin{align}\lab{ES1}
2u=\3F^*\cdot\3F=|\3F|^2&& 2i\3S=\3F^*\times\3F && 2\5I=|\3F^2|.
\end{align}
The combinations $\3E\pm i\3B$ have been called \sl Riemann-Silberstein vectors \rm \ci{B96} and \sl Faraday vectors \rm \ci{B99}. They have been rediscovered many times and were explored extensively by Bateman \ci{B15}. See also \ci[Chapter 9]{K94} and \ci{K4}.

By \eq{I},  
\begin{align}\lab{pos1}
\5I\0x=0\iff \3E\0x^2-\3B\0x^2=\3E\0x\cdot\3B\0x=0\iff \3F^2=0.
\end{align}
An electromagnetic field with $\3F\0x^2=0$ is said to be \sl null at $x$. \rm Nullity is a local, Lorentz-invariant concept. It is the \sl field \rm counterpart of a \sl massless particle. \rm Indeed, 
\begin{align}\lab{vc}
|\3v\0x|=1\iff \5I\0x=0.
\end{align}
\sl Electromagnetic energy propagates at the speed of light only at events $x$ where the field is null. Elsewhere, it travels at speeds less than $c$ and has a positive inertia density. \rm

Although this simple fact should be widely known in classical electrodynamics, I've been unable to find any clear reference to it in the mainstream literature and in discussions with several knowledgeable colleagues. The sole exception, to my knowledge, is a brief note in \ci[page 6]{B15}.\footnote{In \sl quantum \rm electrodynamics it is well known that exchanged photons are \sl virtual \rm (off the zero-mass shell in Fourier space), hence they can have \sl any \rm speed -- possibly greater than $c$. It should be interesting to connect these classical and quantum ideas. They are in some sense \sl dual: \rm  one is local in 4D spacetime, and the other is local in 4D Fourier space.
}

How are we to understand that electromagnetic waves, which are the ingredients of light itself, do not generally travel at the speed of light? The explanation is simple.

\bul  It is well known that every electromagnetic field becomes \sl asymptotically \rm null in the \sl far zone, \rm where its wavefronts become asymptotic to plane waves. But fields are generally not null in the \sl near zone, \rm close to their sources. Hence a \sl generic \rm electromagnetic field (\ie one not chosen in a special way) has a non-vanishing inertia density in the near zone which decays to zero in the far zone. This explains why observed light (which is usually seen in the far zone due to its very short wavelength) is seen to propagate at a speed very close to $c$.

\begin{figure}[ht]% htbp
\begin{center}
\includegraphics[width=4 in]{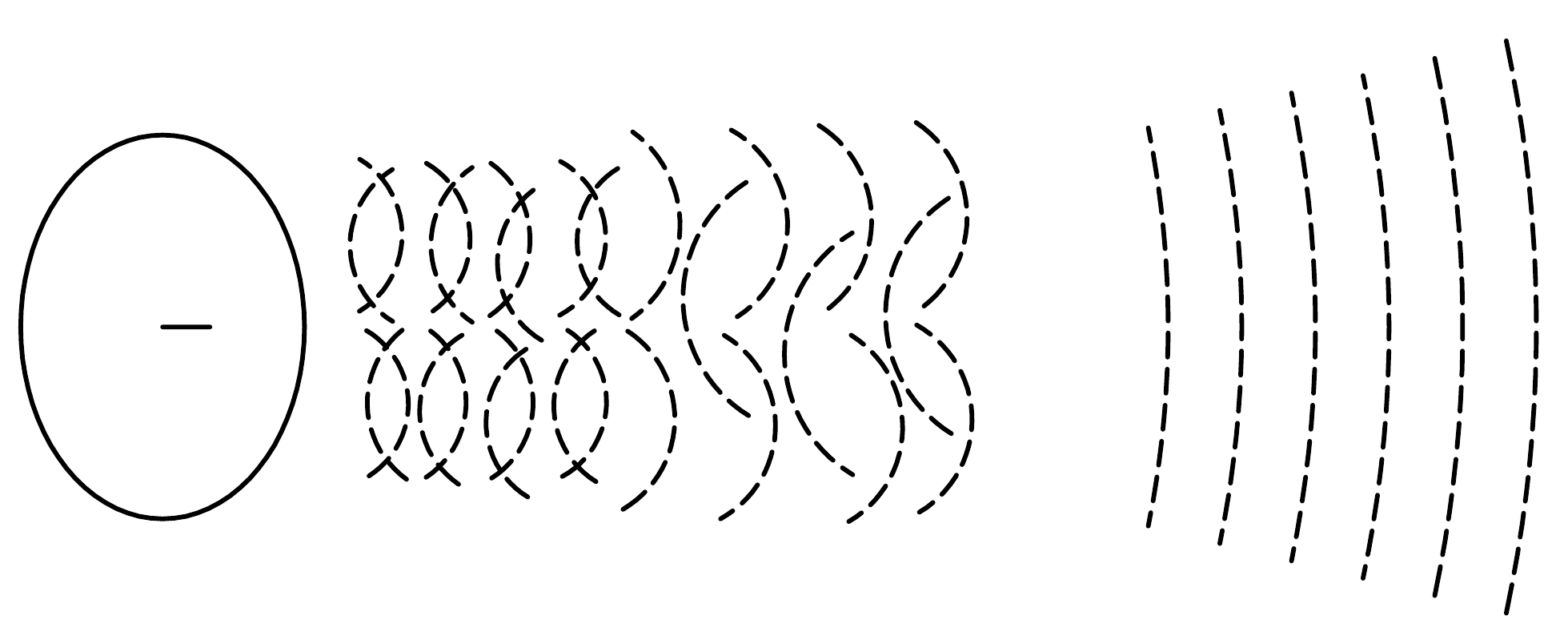}
\caption{\small In the near zone, the energy in a generic wave propagates back to the source as well as away from it.  This \sl incoherence \rm leads to reactive energy and an energy flow speed $v<c$, hence inertia. In the far zone, only the outgoing waves survive and the field becomes asymptotically coherent.}
\label{F:Fig_Near}
\end{center}
\end{figure}

\bul The inertia in the near zone can be understood as the result of \sl interference \rm between different parts of the field. As illustrated in Figure \ref{F:Fig_Near}, parts of the near field propagate away from the source while other parts propagate back toward the source. That causes \sl reactive energy \rm to circulate back and forth between the source and the near field, giving rise to inertia. This has important consequences in antenna theory, where electromagnetic fields tend to loiter near the antenna to a larger or lesser extent. Antennas generating a great deal of reactive energy are inefficient radiators since the associated inertia tends to slow down their transmission time, effectively limiting the bandwidth. For a careful analysis of the reactive energy in the near zone of antennas, see \ci{Y96}. See also \ci{Wik-NF} for a less rigorous but easily comprehensible description.

\bul An important aspect of nullity was discovered by Iwo Bialynicki-Birula \ci{B3}. A generic
electromagnetic field in free space is null along a set of 2-dimensional hypersurfaces $\5S$ in spacetime since $\3F\xt^2=0$ imposes two real conditions on the four spacetime variables $\xt$. The time-slices $\5S_t$ of $\5S$ are (generically) curves in space, and these curves evolve with $t$. Bialynicki-Birula showed that the non-null field surrounding such a curve circulates around it, forming a \sl vortex line \rm along $\5S_t$. The above argument shows that when a field is not null in an extended region of spacetime, it travels at the speed of light only along such \sl electromagnetic vortices. \rm Elsewhere it travels at speeds less than $c$, although the energy flow speed everywhere approaches $c$ in the far zone.

\bul The characteristic property of fields that are null in an \sl extended \rm region of spacetime (not merely on isolated vortex lines, as in the case of a generic field) is that they are \sl coherent, \rm \ie their various parts travel in unison and without interference. They are rewarded by 
having zero inertia and propagating at the exact speed $c$ within the region of coherence.

Various classes of \sl globally \rm null electromagnetic fields (where $\3F^2$ vanishes \sl almost everywhere, \rm with the possible exception of singularities whose supports have zero measure) are known; for examples, see \ci{B15,R61,T62, RT64}.  However, no solutions with \sl extended sources \rm appear to be known which radiate  null fields everywhere outside the source region. 

In this paper we construct a class of localized pulsed-beam fields whose charge-current densities live on a \sl disk \rm in space and which are null everywhere outside the disk.\footnote{More accurately, the fields are also singular on the line forming the \sl axis \rm of the disk, which acts as a  \sl vortex jet. \rm See section \ref{S:coherent}.
}
We call these fields \sl coherent electromagnetic wavelets. \rm If the singular sources radiating such wavelets can be approximated in practice, they should be very efficient radiators.

\section{The scalar pulsed-beam wavelets $\2\Y\xt$}\label{S:scalar}

We now summarize some results regarding pulsed-beam wavelet solutions of the scalar wave equation. Detailed explanations can be found in \ci{K3,HK9}.

Consider the wave $\Y\xt$ radiated by the point source $4\p\d\ox$ driven by a pulse $g_0\0t$, which satisfies the wave equation
\begin{align}\lab{Y0}
\Box\Y\xt=4\p g_0\0t\d\ox,\ \ \hbox{where}\ \  \Box=\pl_t^2-\D
\end{align}
is the wave operator. The unique causal (retarded) solution is 
\begin{align}\lab{Y1}
\Y\xt=\frac{g_0(t-r)}r\=\frac{g_0(t-r/c)}r.
\end{align}
The corresponding \sl scalar pulsed-beam wavelet \rm $\2\Y$ is an extension of $\Y$ to complex spacetime, obtained formally by displacing the point source from the origin to an imaginary point $i\3a$. It is defined by
\begin{align}\lab{Y}
\2\Y\xt=\frac{g(\t-\z)}{\z}
\end{align}
where $\t=t-is$ is a complex time,
\begin{align}\lab{z}
\z=\sr{(\3x-i\3a)^2}=\sr{r^2-a^2-2i\3a\cdot\3x},\qq r=|\3x|,\ a=|\3a|
\end{align}
is the \sl complex distance\rm\footnote{The idea of transforming a spherically symmetric solution of a partial differential equation with a point source to a cylindrically symmetric one by an imaginary displacement of the point source, and the ensuing complex distance, was introduced by Ted Newman in general relativity in 1965. Newman and his collaborators used it to derive the Kerr solution (neutral spinning black hole) and the Kerr-Newman solution (charged spinning black hole) from the spherically symmetric Schwarzschild and ReissnerÐNordstr\"om solutions, respectively \ci{NJ65, N65}. Deschamps discovered the idea independently in 1971 in the context of the Helmholtz equation, where he use it to derive the time-harmonic \sl complex-source beams \rm \ci{D71} from the spherically symmetric Green function, as in \eq{csb}:
\begin{align*}
\frac{e^{ikr}}r\to\frac{e^{ik\z}}{\z}.
\end{align*}
Probably the \sl earliest \rm use of $\z$ was in 1897, when Appell used it to study the analytically continued Newtonian potential $1/\z$ as a harmonic function  \ci{A87}.
}
from the imaginary source point $i\3a$ to the real observation point $\3x$, and $g$ is the \sl analytic signal \rm or positive-frequency part of the driving signal $g_0\0t$, \ie
\begin{align}\lab{as}
g(\t)&=\frac1{2\p}\int_0^\infty\dd\o\, e^{-i\o\t}\1g_0\0\o
=\frac1{2\p}\int_0^\infty\dd\o\, e^{-i\o t}e^{-\o s}\1g_0\0\o.
\end{align}
Here $\1g_0\0\o$ is the usual Fourier transform. If $g_0$ is square-integrable, then $g$ is analytic in the lower-half complex time plane $s>0$ due to the smoothing factor $e^{-\o s}$ in \eq{as}. If $\1g_0\0\o$ decays rapidly as $\o\to\infty$, then $g(t-is)$ may also extend analytically into the upper half-plane. 
For example, in \ci{HK9} we used the Gaussian pulse of \sl duration \rm $d>0$,
\begin{align}\lab{gauss}
g_0\0t=\frac{e^{-t^2/d^2}}{\sr{\p} d},
\end{align}
so that 
\begin{align*}
d\to0\imp \Y\xt\to\frac{\d(t-r)}r.
\end{align*}
This is the \sl retarded wave propagator, \rm a fundamental solution of the wave equation:
\begin{align*}
\Box\Y\xt=4\p\d\0t\d\ox.
\end{align*}
The associated pulsed beams $\2\Y\xt$ were shown to form a `basis' for all radiated solutions of the wave equation by extending Huygens' principle to complex spacetime. Thus $\2\Y$ are `wavelets' in two different senses:
\begin{itemize}
\item From a \sl classical \rm point of view, $\2\Y$ and its spacetime translates $\2\Y(x-x_0)$ are pulsed-beam versions of  \sl Huygens wavelets. \rm The analytic continuation to complex spacetime breaks the spherical symmetry of the latter, deforming them to pulsed beams.

\item From a \sl modern \rm point of view, they are a family of localized functions paramaterized by their centers and widths which can be used to expand a space of functions, as in \sl time-scale analysis \rm \ci{K94}. In this case, the centers are the real spacetime translation parameters $x_0$, the widths are the imaginary translation parameters $(\3a,s)$, and the space of functions represented as superpositions of $\2\Y$ are radiated waves.
\end{itemize}

Returning to the general wavelet $\2\Y$ in \eq{Y}, we \sl define \rm its source $\2\vr$ by
\begin{align}\lab{rho}
4\p\2\vr\xt=\Box\2\Y\xt,
\end{align}
where the wave operator $\Box$ must act in the \sl distributional \rm sense to accommodate the singularities of $\2\Y$. To understand $\2\vr$, note that the complex square root $\z$ has branch points whenever its argument belongs to the negative real axis:
\begin{align*}
r^2-a^2-2i\3a\cdot\3x\le 0\iff r\le a\ \ \hbox{and}\ \ \3a\cdot\3x=0.
\end{align*}
Since we are keeping the imaginary spacetime variables $(\3a,s)$ fixed, the branch points of $\z$ form the \sl disk \rm in $\rr3$ given by
\begin{align}\lab{D}
\5D=\{\3x: r\le a,\ \3a\cdot\3x=0\}=\{\3x: \re\z=0\}.
\end{align}
\sl The formal displacement of the point source from the origin to $i\3a$ transforms the point singularity of $\Y$ to a disk singularity of $\2\Y$ on $\5D$. \rm $\2\Y$ is infinite on the boundary
\begin{align}\lab{C}
\5C=\pl\5D=\{\3x: r=a,\ \3a\cdot\3x=0\}=\{\3x: \z=0\}
\end{align}
and discontinuous across the interior of $\5D$, where $\im\z$ has a jump discontinuity. 
We choose the branch of $\z$ with $\re\z\ge 0$, so that $\z$ is an extension of the positive Euclidean distance, \ie $\z\to r$ as $\3a\to\30$. It then follows that $\2\Y$ is an extension of $\Y$ in the sense that (a) its source $\2\vr\xt$ vanishes whenever $\3x\notin\5D$, and (b)
\begin{align*}
\3a\to\30,\ s\to0 \imp 2\re\2\Y\xt\to\frac{g_0(t-r)}r\= \Y\xt.
\end{align*}
Taking the real part is necessary to restore the negative-frequency component of $g_0$.
The distribution $\2\vr$ in \eq{rho} has been computed rigorously in \ci{K0,K4a}.

\begin{figure}[ht]% htbp
\begin{center}
\includegraphics[width=3 in]{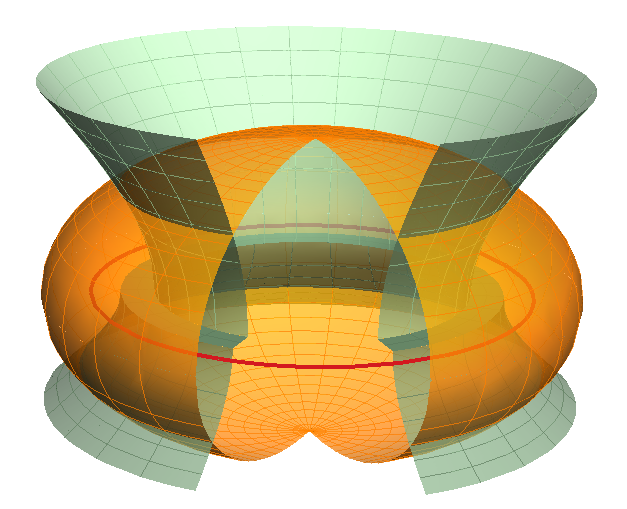}
\caption{\small The real and imaginary parts of $\z=\x-i\h$, together with the azimuthal angle $\f$, form an oblate spheroidal coordinate system $(\x,\h,\f)$ in $\rr3$. The above plot shows  cut-away views of an oblate spheroid $\5O_\x$ with $\x=0.7a$, a semi-hyperboloid $\5H_\h$ with $\h=0.8a\, (z>0)$ and another with $\h=-0.5a \,(z<0)$. Also shown is the focal circle $\5C$ with radius $a$ at $z=0$, whose interior is the branch disk $\5D$ of $\z$.}
\label{F:Fig_OSCS_Color}
\end{center}
\end{figure}

As its name indicates, $\2\Y\xt$ is a pulsed beam radiated by the disk $\5D$ and propagating in the direction of $\3a$. To see this, write
\begin{align}\lab{z1}
\z=\x-i\h,\qq \x\ge 0.
\end{align}
Choosing a cylindrical coordinate system $(\r,\f,z)$ with $z$-axis parallel to $\3a$, we have
\begin{align*}
\z^2=(\3x-i\3a)^2=r^2-a^2-2i\3a\cdot\3x=\x^2-\h^2-2i\x\h,
\end{align*}
hence
\begin{align*}
r^2-a^2=\x^2-\h^2\ \ \hbox{and}\ \ az=\x\h
\end{align*}
and
\begin{align*}
a^2\r^2=a^2r^2-a^2z^2=a^2(a^2+\x^2-\h^2)-\x^2\h^2.
\end{align*}
Therefore $(\x,\h)$ are related to $(\r,z)$ as follows:
\begin{align}\lab{rz}
\bx{a^2\r^2=(a^2+\x^2)(a^2-\h^2)\ \ \hbox{and}\ \ az=\x\h.}
\end{align}
In particular, it follows that
\begin{itemize}
\item $0\le \x\le r $ and $-a\le\h\le a$,
\item the level surfaces of $\x$ are the \sl oblate spheroids \rm
\begin{align}\lab{Ox}
\5O_\x=\LB\3x: \frac{\r^2}{a^2+\x^2}+\frac{z^2}{\x^2}=1\RB,\qq \x>0,
\end{align}
\item the level surfaces of $\h$ are  the \sl semi-hyperboloids \rm
\begin{align}\lab{Hh}
\5H_\h=\LB\3x: \frac{\r^2}{a^2-\h^2}-\frac{z^2}{\h^2}=1,\ \sgn z=\sgn\h\RB,\qq 0<\h^2<a^2.
\end{align}
\end{itemize}
The \sl far zone \rm consists of points $\3x$ with $r\gg a$, where the observer is far from the disk. By \eq{z} and \eq{z1},
\begin{align}\lab{far}
r\gg a\imp \x\app r\ \ \hbox{and}\ \ \h\app\3a\cdot\bh x\=a\cos\q,\qq \bh x=\3x/r.
\end{align}
Hence $\x$ is a deformation of $r$ and $\h$ is a deformation of $a\cos\q$. For $\x\gg a$, $\5O_\x$ becomes asymptotic to the sphere of radius $r=\x$ and $\5H_\h$ becomes asymptotic to the cone $\cos\q=\h/a$. The circle $\5C$ \eq{C} is the common  \sl focal set \rm of all the spheroids and hyperboloids. See Figure \ref{F:Fig_OSCS_Color}.

To see that $\2\Y\xt$ is a pulsed beam propagating in the direction of $\3a$, let $g_0\0t$ be a pulse peaking around $t=0$. Then at a fixed observation point $\3x$,
\begin{align}\lab{g}
g(\t-\z)=\frac1{2\p}\int_0^\infty\dd\o\,e^{-i\o(t-\x)}e^{-\o(s-\h)}\1g_0\0\o
\end{align}
is a pulse peaking around $t=\x\ox$. Hence the \sl wavefronts \rm of $\2\Y\xt$ consist of the oblate spheroids $\5O_\x$, just as the wavefronts of $\Y\xt$ in \eq{Y1} are the spheres $r=t$.

But the pulse amplitude is not uniform on $\5O_\x$. Due to the factor $e^{-\o(s-\h)}$ in \eq{g}, it is strongest along the positive $z$-axis, where $\h=a$, and weakest along the negative $z$-axis, where $\h=-a$. In fact, if we follow a propagating wavefront along any given semi-hyperboloid $\5H_\h$ (\ie fix $\h=\h_0$ and let $\x$ increase as $\x\0t=t+\x_0$), then
\begin{align*}
\2\Y\xt=\frac{g(-\x_0-i(s-\h_0))}{t+\x_0-i\h_0}\sim\frac{g(-\x_0-i(s-\h_0))}t,\qq t\gg |\x_0-i\h_0|.
\end{align*}
The numerator remains constant, and the attenuation of $\2\Y$ is due entirely to the increasing distance, \ie size of the denominator. This shows that $\2\Y$ propagates along $\5H_\h$, so each $\5H_\h$ is a \sl diffraction surface \rm for the pulsed beam. In the far zone, $\5O_\x$ becomes a spherical wavefront ($\x\app r$) and $\5H_\h$ becomes a \sl diffraction cone \rm ($\h\app a\cos\q$) for the beam.

In the frequency domain, $\2\Y$ is represented by
\begin{align}\lab{csb}
\2\Y_\o\ox=\1g_0\0\o\frac{e^{ik\z}}{\z}=\1g_0\0\o\frac{e^{ik\x}}{\z}\,e^{k\h},\qq  \o>0,
\end{align}
where $k=\o/c\=\o$ is the wave number and we have suppressed the factor $e^{-\o s}$, which controls the growth as $\o\to\infty$. These are the original \sl complex-source beams \rm derived by Deschamps \ci{D71}. In the far zone,
\begin{align*}
r\gg a\imp \2\Y_\o\ox \sim\1g_0\0\o\frac{e^{ikr}}r\,e^{ka\cos\q},
\end{align*}
showing that the radiation pattern of the beam is 
\begin{align*}
\5F_\o\0\q=\1g_0\0\o\,e^{ka\cos\q}.
\end{align*}
Applications of complex-source beams in electrical engineering were developed extensively by Felsen and later extended to the time domain by Heyman et al. using analytic signals. See \ci{F76,HF89}; also \ci{HK9} and references therein.

The \sl spacetime analytic-signal transform, \rm  a tool for extending solutions of wave equations (including the Klein-Gordon and Dirac equations in quantum mechanics) to complex spacetime and their interpretation as directed wave packets (relativistic coherent states), was introduced and developed in \ci{K77, K78, K87, K90, K94, K3}.

\bf Remark. \rm The analytic signal $g\0\t$ can be interpreted as follows. Whereas the point source  $\d\ox$ can be driven by an arbitrary time signal $g_0\0t$, the disk $\5D$ must be driven by a separate time signal $g_0(\3\r,t)$ for each $\3\r\in\5D$. However, these signals cannot be independent since the points $\3\r$ are rigidly connected to $\5D$. By the magic of complex analysis, the single analytic function $g(t-is)$ represents a \sl coherent \rm driving signal for the entire disk. The imaginary time parameter $s$ represents a \sl reaction time \rm measuring the degree of correlation. Increasing $s$ makes $g(t-is)$ smoother, hence more correlated. In \eq{g}, for example, we must require that
\begin{align*}
s-\h\ge 0\ \forall \3x,\ \ \hbox{hence}\ \  s\ge a\=a/c,
\end{align*}
in order for the integral  to converge when $g$ is a general square-integrable signal. This implies that the correlations between all points on $\5D$ are related \sl  causally. \rm

\section{The twisted null congruence of $\2\Y$}\label{S:congr}

We now show that the oblate spheroids $\5O_\x$ and semi-hyperboloids $\5H_\h$ can be understood  \sl geometrically \rm as the wavefronts and diffraction surfaces, respectively, of `light rays' (points moving in straight lines at speed $c$) emitted from the disk $\5D$ if the latter is assumed to \sl spin \rm at the extreme relativistic angular velocity $c/a$. This will later inspire the construction of \sl coherent electromagnetic wavelets, \rm which fully justify the interpretation in terms of light rays.

We have shown that $\5H_\h$ should be interpreted as a diffraction surface for the pulsed beam $\2\Y$, \ie a surface along which $\z\2\Y=g(\t-\z)$ is constant when the wavefront $\5O_\x$ expands as $\x=t$. We now refine this interpretation.

In Figure \ref{F:Fig_OSCS_Color}, imagine that $\5H_\h$ consists of individual `rays' propagating away from the source disk in the direction orthogonal to the wavefront $\5O_\x$. The location of the points tracing out the rays is given in oblate spheroidal coordinates $(\x,\h,\f)$ with 
\begin{align*}
\x=t,\qq \h=\h_0,\qq \f=\f_0,
\end{align*}
with $\h_0$ and $\f_0$ fixed. We shall use $\x$ as the time parameter to describe the motion.
The trajectory of the ray in space is then
\begin{align}\lab{xt}
\3x_0\0\x=(\r\0\x\cos\f_0,\r\0\x\sin\f_0,z\0\x),
\end{align}
and \eq{rz} gives
\begin{align*}
a^2\r\0\x^2=(\x^2+a^2)(a^2-\h_0^2),\qq az\0\x=\x\h_0\,.
\end{align*}
Its velocity is
\begin{align*}
\3v_0\0\x=\pl_t\3x_0\0\x=(\r_\x\cos\f,\r_\x\sin\f,z_\x)
\end{align*}
where
\begin{align*}
\r_\x\=\pl_\x\r=\frac{\x(a^2-\h_0^2)}{a^2\r},\qq z_\x\=\pl_\x z=\frac{\h_0}a.
\end{align*}
It follows that
\begin{align}\lab{v0}
\3v_0^2=\r_\x^2+z_\x^2=\frac{\x^2+\h_0^2}{\x^2+a^2}\,.
\end{align}
Since $\r\00^2=a^2-\h_0^2$, the rays starting from the center of the disk ($\r\00=0$) have $\h_0=\pm a$ and travel in straight lines with speed $|\3v_0|=1$. But all other rays, starting with $0<\r\00\le a$, have $\h_0^2=a^2-\r\00^2<a^2$, hence their propagation speed is $|\3v_0|<1$ and their paths are hyperbolic.

Ideally, we would like our `rays' to travel like light in free space: in straight lines, and at speed $c$.
This can be arranged by allowing $\f$ to vary in time. Let the ray have the trajectory
\begin{align*}
\3x\0\x=(\r\0\x\cos\f\0\x,\r\0\x\sin\f\0\x,z\0\x),
\end{align*}
so that its velocity is
\begin{align*}
\3v\0\x\=\pl_t\3x\0\x=\3v_0\0\x+\r\0\x\f'\0\x\bh\f
\end{align*}
where
\begin{align*}
\bh\f=\pl_\f\bh\r=\pl_\f(\cos\f,\sin\f,0)=(-\sin\f,\cos\f,0).
\end{align*}
We require that $|\3v\0\x|\=1$, hence \eq{v0} gives
\begin{align*}
\r^2\f'^2=1-\frac{\x^2+\h_0^2}{\x^2+a^2}=\frac{a^2-\h_0^2}{\x^2+a^2}
\imp \f'^2=\frac{a^2}{(\x^2+a^2)^2}.
\end{align*}
That is, \sl the points of the circle
\begin{align*}
\5C_{\x-i\h_0}=\5O_\x\cap\5H_{\h_0}
\end{align*}
propagate precisely at the speed $c=1$ if and only if $\5C_{\x-i\h_0}$ spins at the angular velocity \rm
\begin{align}\lab{opm}
\bx{\o_\pm\0\x=\f'\0\x=\pm\frac{a}{\x^2+a^2}\=\pm\frac{ca}{\x^2+a^2}.}
\end{align}
Furthermore, since $\f'\0\x$ is independent of $\h_0$, \sl all \rm the circles $\5C_{\x-i\h_0}$ with given $\x$ spin at the same angular velocity and \sl the oblate spheroid $\5O_\x$ spins as a rigid body. \rm  We say that the spheroid with angular velocity $\o_+$ has \sl positive helicity \rm and that with  angular velocity $\o_-$ has \sl negative helicity. \rm 

Since the branch disk $\5D$ is the limit of $\5O_\x$ as $\x\to 0$, it follows that $\5D$ spins rigidly at $\o_\pm\00=\pm c/a$. This is precisely the angular velocity at which the points on the boundary $\pl\5D$ move at the speed of light! More will be said about this in Section \ref{S:Newman}.

Let us now compute the trajectory of a single point of  $\5C_{\x-i\h_0}$.  We can set $\f_0=0$ without loss of generality. Integrating \eq{opm} gives
\begin{align}\lab{fpm}
\f\0\x=\pm\tan\inv(\x/a),
\end{align}
hence
\begin{align*}
\cos\f\0\x=\frac a{\sr{\x^2+a^2}}\ \ \hbox{and}\ \ \sin\f\0\x=\pm\frac\x{\sr{\x^2+a^2}}.
\end{align*}
Furthermore, $a\r\0\x=\sr{(\x^2+a^2)(a^2-\h_0^2)}$, so
\begin{align*}
\r\0\x\cos\f\0\x&=\sr{a^2-\h_0^2}=\r_0,\qq
\r\0\x\sin\f\0\x=\pm\frac{\x\r_0}a,\qq
z\0\x=\frac{t\h_0}a.
\end{align*}
The trajectory of the ray starting at the point on $\5D$ with $\r=\r_0$ and $\f=0$ is therefore
\begin{align}\lab{xx}
\3x\0\x=\lp\r_0, \pm\frac{\x\r_0}a, \frac{\x\h_0}a\rp=\3x\00+\3v_\pm \x,
\end{align}
where the velocity is given in Cartesian coordinates by
\begin{align*}
\3v_\pm=a\inv\lp 0, \pm\r_0, \h_0\rp,\qq \3v_\pm^2=1.
\end{align*}
Since the choice $\f\00=0$ is arbitrary, we have proved the following result.

{\thm Every point emanating from $\5D$ which follows the exploding wavefront $\5O_\x$ with $\x=ct$ as it spins at the angular velocity $\o_\pm\0\x$ while constrained to $\5H_{\h_0}$ moves in a straight line at speed $c$. 
}\label{T:OxHh}

The trajectories of such points might be interpreted as `light rays,' except that they are associated with a solution of the scalar wave equation rather than Maxwell's equations. This will be corrected later.

\begin{figure}[ht]% htbp
\begin{center}
\includegraphics[width=4 in]{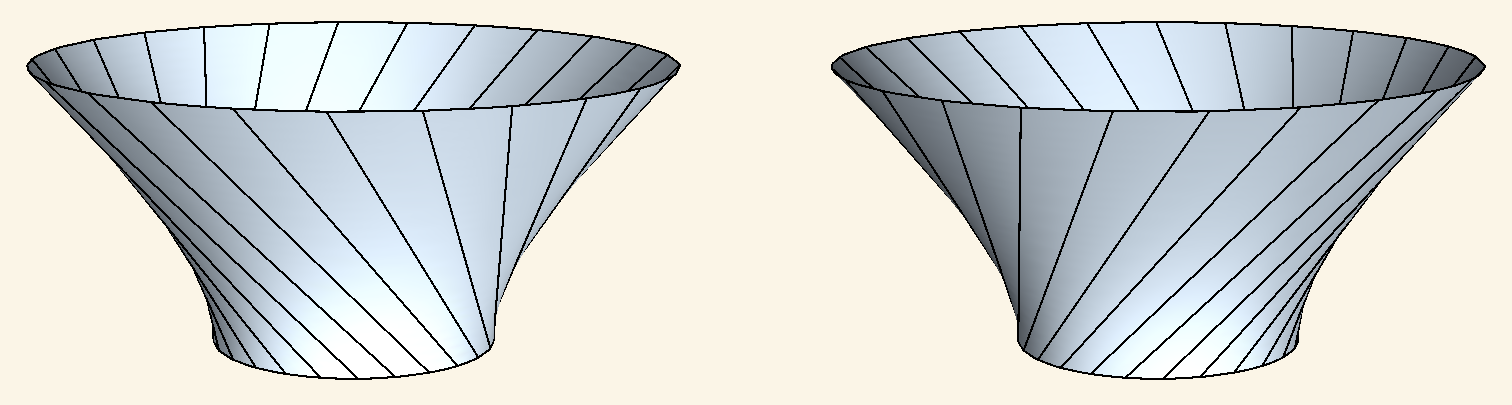}
\caption{\small The semi-hyperboloids $\5H_\h$ are  \sl doubly ruled surfaces, \rm with the straight lines interpreted as `light rays' radiated by $\5D$ and traveling at speed $c$.}
\label{F:Fig_Cong}
\end{center}
\end{figure}

The existence of two families of straight lines generating $\5H_\h$ is a well-known fact of geometry: every circular hyperboloid of one sheet is a \sl doubly ruled surface. \rm 
Our \sl semi\rm-hyperboloids are ruled by the rays, which are half-lines emanating from a circle on $\5D$ as shown in Figure \ref{F:Fig_Cong}.  The rays move upwards if $\h>0$ and downwards if $\h<0$.\footnote{In our case, the ruling for $\h<0$ has the \sl opposite \rm helicity from that for $\h>0$. That is because when the disk spins counterclockwise as seen from the top, it spins clockwise as seen from the bottom.
}
The motion is vertical if $\h=\pm a\  (\r=0)$. It acquires a horizontal component when $0<|\h|<a$, and becomes purely horizontal as $\h\to0\ (\r\to a)$. For $\h=0$, the trajectory begins on the branch circle $\5C$ and is \sl tangent \rm to $\5C$, remaining in the $xy$-plane. That is so because $\5C$ is already moving at the speed of light, so the vertical velocity component must vanish. The union of all such horizontal rays forms the degenerate semi-hyperboloid
\begin{align}\lab{H0}
\5H_0=\{\3x: \h=0\}=\{\3x: r\ge a, z=0\}.
\end{align}

Let us now compute the velocity vector of the ray going through an \sl arbitrary \rm point $\3x\in\rr3$. Identifying $t$ again with $\x$, define
\begin{align*}
\3u\ox&\=\frac{\pl\3x}{\pl\x}=\frac{\pl\3x}{\pl\r}\frac{\pl\r}{\pl\x}
+\frac{\pl\3x}{\pl z}\frac{\pl z}{\pl\x}+\frac{\pl\3x}{\pl\f}\frac{\pl\f}{\pl\x}.
\end{align*}
Using \eq{rz} and
\begin{align*}
\frac{\pl\3x}{\pl\r}=\bh\r &&\frac{\pl\3x}{\pl z}=\bh z && \frac{\pl\3x}{\pl\f}=\r\bh\f,
\end{align*}
we obtain
\begin{align}\lab{upm}
\bx{\3u_\pm\ox=\bh\r\,\frac\x a\sr{\frac{a^2-\h^2}{\x^2+a^2}}+\bh z\,\frac\h a\pm\bh\f\sr{\frac{a^2-\h^2}{\x^2+a^2}}.}
\end{align} 
In terms of the orthonormal oblate spheroidal basis $(\bh\x, \bh\h,\bh\f)$, this is seen to be
\begin{align}\lab{upmos}
\3u_\pm\ox=\bh\x\sr{\frac{\x^2+\h^2}{\x^2+a^2}}\pm\bh\f\sr{\frac{a^2-\h^2}{\x^2+a^2}}.
\end{align}
As expected, the coefficient of $\bh\h$ vanishes since the rays propagate with constant $\h$. Furthermore, $\3u_\pm$ is not orthogonal to the wavefront $\5O_\x$ since 
\begin{align*}
\bh\x\times\3u_\pm=\pm\bh\h \sr{\frac{a^2-\h^2}{\x^2+a^2}}.
\end{align*}

Note that the $\f$-component of $\3u_\pm$ does not depend on the sign of $\h$. This confirms that the upper and lower semi-hyperboloids ($\pm\h>0$) are ruled oppositely. Both \eq{upm} and \eq{upmos} show that $|\3u_\pm|=1$, as expected.

\begin{figure}[ht]% htbp
\begin{center}
\includegraphics[width=2.6 in]{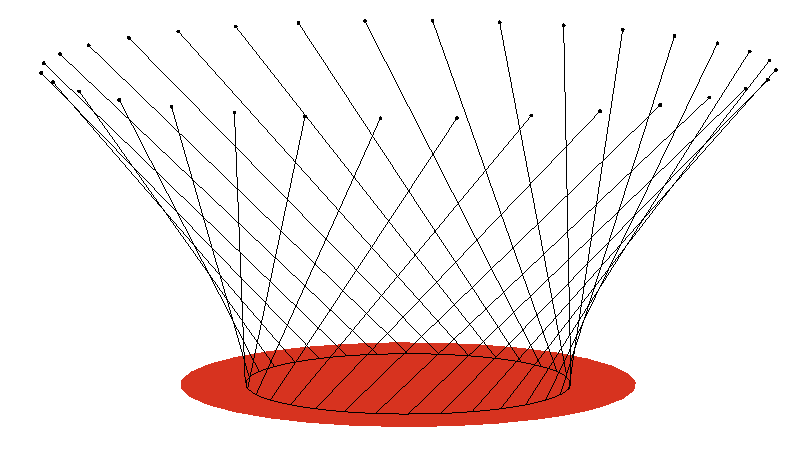}\sh{-2}
\includegraphics[width=2.1 in]{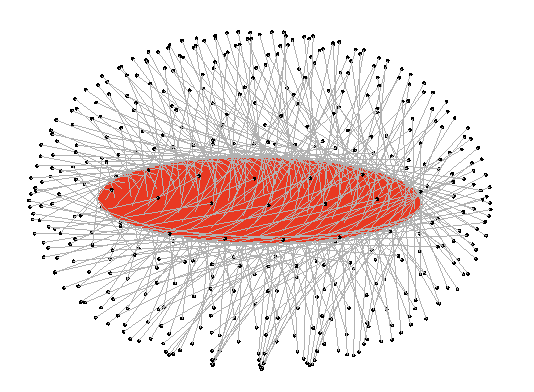}
\caption{\small  This picture gives a geometric explanation of $\5H_\h$ and $\5O_\x$ in terms of the disk $\5D$ spinning at the extreme relativistic angular velocity $c/a$. \sl Left: \rm The envelope of the rays emitted from a circle or radius $0<\r<a$ on $\5D$ is the diffraction semi-hyperboloid $\5H_\h$ with $\h=\pm\sr{a^2-\r^2}$. (We show only $\5H_\h$ with $\h=0.7$.) The larger $\r$, the greater the horizontal speed and the more \sl twisted \rm is the bundle of rays generating $\5H_\h$. \sl Right: \rm The `photons' emitted from the entire disk $\5D$ at $t=0$ form the wavefront $\5O_\x$ at $t=\x$.}
\label{F:Fig_spin}
\end{center}
\end{figure}

The above picture of radiation from a rotating disk explains both the wavefronts $\5O_\x$ and the diffraction surfaces $\5H_\h$. With some poetic license, let us call the emitted particles `photons' even though $\2\Y$ is a scalar potential and not an electromagnetic potential.\footnote{In Section \ref{S:EM} we shall reinterpret $\2\Y$ as the scalar component of an electromagnetic 4-vector potential, so that the terminology, while still classical, is not entirely inappropriate.
}
As shown above, the envelope of all the rays emanating from a fixed circle of radius $\r_0<a$ in $\5D$ is the pair of semi-hyperboloids $\5H_{\pm\h_0}$ with $\h_0=\sr{a^2-\r_0^2}$. However, if we take a snapshot of \sl all \rm the photons at fixed time $t=\x>0$, we find that they form the wavefront $\5O_\x$. This is illustrated in Figure \ref{F:Fig_spin}. The radiation from a spinning disk thus explains the level surfaces of both $\x$ and $\h$, showing the complex distance $\z=\x-i\h$ is a natural tool for its analysis. 

It can be shown that the \sl vorticity \rm of $\3u_\pm$ is given by
\begin{align}\lab{twist}
\3\o_\pm\=\curl\3u_\pm=\pm\frac{2\h}{\x^2+\h^2}\,\3u_\pm.
\end{align}
Thus $\3u_\pm$ is a \sl Beltrami field, \rm curling around its own axis with helicity $\pm$. 

There is a natural four-dimensional (spacetime) formulation of the above. Consider the two stationary 4-vector fields $u_\pm\ox$ with contravariant and covariant components
\begin{align}\lab{congr}
u_\pm^\m\ox=(\3u_\pm\ox,1)\qq u^\pm_\m\ox=(-\3u_\pm\ox,1),\qq \m=1,2,3,4.
\end{align}
They are \sl null, \rm \ie their Lorentzian squares with respect to the Minkowski metric $\h_{\m\n}={\rm diag}\,(-1,-1,-1, 1)$ vanish:
\begin{align*}
u^2_\pm\=\h_{\m\n}u_\pm^\n u_\pm^\m=
u^\pm_\m u_\pm^\m=1-\3u_\pm^2=0,
\end{align*}
where we sum as usual over the repeated index $\m$. $u_\pm$ are the 4-velocities of the twisted ray bundles with helicity $\pm$, and they determine the rays by\footnote{Here $t$ ranges over all real numbers and there is a unique $t_0$ for which $\3x+t_0\3u_\pm\in\5D$. For $t>t_0$, $x_\pm\xt$ has helicity $\pm$, while for $t<t_0$ it has helicity $\mp$.
}
\begin{align}\lab{congr0}
x_\pm^\m\xt=(\3x,0)+t u_\pm^\m\ox=(\3x+t\3u_\pm\ox, t).
\end{align}
In general relativity, ray bundles (possibly in curved spacetime) such as \eq{congr0} are called \sl null congruences, \rm  and they play an important role in the construction and analysis of nontrivial solutions to the Einstein and Einstein-Maxwell equations; see Section \ref{S:KN}. Note that our rays are not orthogonal to the wavefronts $\5O_\x$. This is a characteristic of \sl twisted \rm congruences, which are characterized by $\3u\cdot\curl\3u\ne 0$.\footnote{In general, a four-dimensional 
version of \eq{twist1} must hold. But since $\3u_\pm$ depends only on $\3x$ and the time component $u_\pm^0=1$, this reduces to \eq{twist1}.
}
In our case,
\begin{align}\lab{twist1}
\3u_\pm\cdot\curl\3u_\pm=\pm\frac{2\h}{\x^2+\h^2}\ne 0\qq\forall z\ne 0.
\end{align}
The congruence is twisting everywhere outside the $xy$-plane, where its rays span the degenerate hyperboloid $\5H_0$ \eq{H0} of half-lines tangent to $\5C=\pl\5D$.

\bf Remark. \rm It will be shown in Section \ref{S:KN} that the null congruence $u^\m_\pm$ in \eq{congr} is \sl identical \rm to the null congruences of the Kerr and Kerr-Newman metrics in general relativity  \ci{R61,T62,RT64,NJ65,N65}. This is an intriguing fact whose significance remains to be fully understood since the Kerr metric and electromagnetic fields are stationary while $\2\Y$ is time-dependent.

\section{Newman's `magic' electromagnetic field}\label{S:Newman}

In 1973, Ted Newman published a paper \ci{N73} detailing the electromagnetic part of the Kerr-Newman solution of the Einstein-Maxwell equations in general relativity \ci{N65}, which represents a spinning, charged black hole. Although the Kerr-Newman solution is quite complicated, its purely electromagnetic part is a simple static field in flat spacetime. Newman begins with the analytically continued Coulomb field obtained by replacing the Coulomb potential $r\inv$ by $\z\inv$:
\begin{align}\lab{Newman}
\bt E\ox=-\grad\frac1{\z}=\frac{\3x-i\3a}{\z^3}.
\end{align}
He then notes that $\bt E$ has the following multipole structure:
\begin{align}\lab{multi}
\re\bt E&=\hbox{(electric monopole)}+\hbox{(electric quadrupole)}+\cdots\\
\im\bt E&=\hbox{(magnetic dipole)}+\hbox{(magnetic octupole)}+\cdots.\nt
\end{align}
For example, using the far-zone approximation \eq{far} gives
\begin{align*}
r\gg a\imp \bt E&=\frac{\3x-i\3a}{(r-i\bh x\cdot\3a)^3}
\sim\frac{\3x}{r^3}+i\frac{3\bh x\cdot\3a-\3a}{r^3}+\5O(a^2/r^4).
\end{align*}
The first term is a real electric monopole of charge $q=1$, and the second term is an imaginary magnetic dipole of magnetic moment $\3a$. Newman therefore interprets $\bt E$ by writing it as
\begin{align}\lab{Newman1}
\bt E\ox\=\3E\ox+i\3B\ox,
\end{align}
where $\3E$ a real electrostatic field and $\3B$ is a real magnetostatic field.

In \ci{K4b}, I computed the charge- and current densities of the field $(\3E, \3B)$ as the \sl distributions \rm in $\rr3$ defined by
\begin{align}\lab{rJ}
4\p\vr\ox=\div\bt E\ox,\qq 4\p\3J\ox=-i\curl\bt E\ox-\pl_t\bt E\ox=-i\curl\bt E\ox,
\end{align}
which is just the complex form of the usual Maxwell equations. The use of distribution theory is necessary because $\bt E$ is discontinuous on the branch cut $\5D$ of $\z$ and singular on the focal circle $\5C=\pl\5D$, where $\z=0$. I showed that: 

\bul $\vr$ and $\3J$ are \sl real \rm distributions supported on $\5D$, so that the homogeneous Maxwell equations are satisfied and thus no \sl magnetic \rm charges or currents are introduced by the complexification.

\bul The \sl surface \rm charge-current density on the \sl interior \rm of $\5D$ is given by
\begin{align}\lab{surf}
\s=-\frac{2a}{(a^2-\r^2)^{3/2}},\qq
\3K=-\frac{2\r\,\bh\f}{(a^2-\r^2)^{3/2}}.
\end{align}
This is the charge-current density of a \sl rigidly spinning disk \rm rotating in the counterclockwise direction $\bh\f$ with angular velocity\footnote{The reader may wonder how $c$ has entered an expression obtained by analytically continuing the Coulomb field. The answer is that the dimensionally correct expression for the current \eq{rJ} is
\begin{align*}
4\p\3J\ox=-ic\curl\bt E\ox-\pl_t\bt E\ox=-ic\curl\bt E\ox,
\end{align*}
so $\3K$ must be multiplied by $c$.
}
\begin{align}\lab{omega}
\o=\frac{|\3K|}{\s\r}=\frac1a\=\frac ca.
\end{align}
The focal circle $\5C$ therefore spins at the speed of light.\footnote{This bizarre property is related to the fact that $\bt E$ is the flat-spacetime limit of the electromagnetic field around a spinning charged black hole. That also explains the singular nature of $\vr$
 and $\3K$ on $\5C$. See \ci{N65, N73}.
 }
Note that the speed of a point on $\5D$ is $v=c\r/a$, hence \eq{surf} can be rewritten in a form displaying its relativistic nature:
\begin{align}\lab{surf1}
\s=-\frac{2}{a^2(1-v^2/c^2)^{3/2}},\qq
\3K=-\frac{2c\r\,\bh\f}{a^3(1-v^2/c^2)^{3/2}}.
\end{align}
The surface sources \eq{surf} can be obtained easily from the discontinuities of $\3E$ and $\3B$ across the branch cut $\5D$.  As $\3x\to\5D$ from above and below, we have
\begin{align*}
 \r<a,\  z\to 0^\pm\imp\z\to\mp i\sr{a^2-\r^2},\qq\3x\to \r\bh\r\=\3\r.
\end{align*}
The upper and lower boundary values of $\bt E$ are thus
\begin{align*}
\bt E^\pm\to\frac{\3\r-i\3a}{\pm i(a^2-\r^2)^{3/2}},
\end{align*}
showing that
\begin{align*}
\3E^\pm=\mp\frac{\3a}{(a^2-\r^2)^{3/2}},\qq  \3B^\pm=\mp\frac{\3\r}{(a^2-\r^2)^{3/2}},
\end{align*}
and the discontinuities are
\begin{align*}
\d\3E&\=\3E^+-\3E^-=-\frac{2\3a}{(a^2-\r^2)^{3/2}}\\
\d\3B&\=\3B^+-\3B^-=-\frac{2\3\r}{(a^2-\r^2)^{3/2}}.
\end{align*}
The jump conditions at an interface \ci[page 18]{J99} thus give
\begin{align}\lab{surf2}
\s&=\bh z\cdot\d\3E=-\frac{2a}{(a^2-\r^2)^{3/2}}&&
\3K=\bh z\times\d\3B=-\frac{2\r\,\bh\f}{(a^2-\r^2)^{3/2}},
\end{align}
proving \eq{surf}. Furthermore, 
\begin{align}\lab{nomag}
\bh z\cdot\d\3B=0\ \ \hbox{and}\ \ \bh z\times\d\3E=\30,
\end{align}
proving that the \sl magnetic \rm charge-current density on $\5D$ vanishes as required by the homogeneous Maxwell equations.

However, these surface sources do not give the entire story. The jump conditions do not account for a 1-dimensional charge-current distribution along the \sl boundary \rm $\5C=\pl\5D$ since the usual method \ci[page 18]{J99} of infinitesimal pillboxes and rectangular line integrals fails there.  This failure is reflected by the fact that the total charge of $\s$ is $q=-\infty$, whereas the field $\bt E$ was derived under the assumption that $q=1$. 

The solution to this puzzle is to compute the complete \sl volume \rm charge-current densities, defined as \sl distributions \rm using \eq{rJ}. When they are computed rigorously (this requires a \sl regularization\rm), we recover \eq{surf} on the interior of $\5D$ but also find that $\pl\5D$ carries a linear charge-current distribution of charge $+\infty$, precisely so that the total charge remains $q=1$.
 
It is interesting to compute the inertia density $\5I$ and energy flow velocity $\3v$ of the Newman field $\bt E$. By \eq{ES1}, 
\begin{align}\lab{NewI}
\bt E^2=\frac1{\z^4},\ \ \hbox{hence}\ \ \5I=\frac1{2|\z^4|}=\frac1{2(\x^2+\h^2)^2},
\end{align}
which is infinite on the focal circle $\5C$ where $\z=0$. Its energy density and Poynting vector are
\begin{align}\lab{NewS}
u&=\frac12|\bt E|^2=\frac{|\3x-i\3a|^2}{2|\z|^4}=\frac{r^2+a^2}{2(\x^2+\h^2)^2}
=\frac{\x^2-\h^2+2a^2}{2(\x^2+\h^2)^2}\\
\3S&=\frac i2\bt E\times\bt E^*=\frac{\3a\times\3x}{|\z|^4}=\frac{a\r}{(\x^2+\h^2)^2 }\,\bh\f,\nt
\end{align}
so the energy flow velocity is
\begin{align}\lab{Newv}
\3v=\frac{\3S}{u}=\frac{2a\r}{2a^2+\x^2-\h^2}\bh\f.
\end{align}
It follows from \eq{rz} with $X=a^2+\x^2$ and $Y=a^2-\h^2$ that
\begin{align*}
\3v^2&=\frac{4XY}{(X+Y)^2}=1-\frac{(X-Y)^2}{(X+Y)^2}\le 1\=c,
\end{align*}
with equality only when $\x=\h=0$, \ie $\3x\in\5C$.

\sv1
\bf Conclusions: \rm

\bul Note the similarity between the charge flow \eq{omega} on $\5D$ and the twisted congruence derived from $\2\Y$ in Section \ref{S:congr}. In both cases, $\5D$ is seen to spin at the extreme relativistic rate $\o=c/a$. This will be the inspiration for reinterpreting $\2\Y$ as the scalar part of an electromagnetic 4-potential $(\bt A,\2\Y)$ whose field is a \sl pulsed \rm version of the Newman field $\bt E$ (Section \ref{S:EM}).

\bul The energy field circulates about the $z$-axis without propagating out \eq{Newv}, which is expected since the field is non-radiating. Its angular velocity at $\3x$ is
\begin{align}\lab{Omega}
\O\ox=\frac{v}\r=\frac{2a}{2a^2+\x^2-\h^2}.
\end{align}
For fixed $\x>0$ \ie $\3x\in\5O_\x$, $\O\ox$ is a minimum on the equator $(\h=0)$ and increases steadily to its maximum at the north and south poles as $|\h|$ increases from 0 to $a$. Hence \sl the energy flow forms a vortex along the $z$-axis. \rm

\bul On $\5D \ (\x=0)$, we have
\begin{align*}
\O(\3\r)=\frac{2a}{2a^2-\h^2}=\frac{2a}{a^2+\r^2}\=\frac{2ac}{a^2+\r^2},\qq \r\le a.
\end{align*}
The angular velocity of the \sl energy \rm flow on $\5D$ differs from the \sl uniform \rm angular velocity $\o$ \eq{omega} of the \sl charge \rm flow. It increases from its minimum value $\O\0a=c/a=\o$ on $\5C$ to its maximum value $\O\00=2\o$ at the origin, which is the center of the vortex.

\bul The \sl real \rm Coulomb field is stationary, as expected:
\begin{align*}
a\to0\imp\3v\to\30.
\end{align*}
The rotation thus comes directly from the analytic continuation $r\to\z$.

\section{From scalar to electromagnetic wavelets}\label{S:EM}

In this section we introduce the idea of \sl electromagnetic pulsed-beam wavelets \rm as time-dependent (pulsed) versions of Newman's static field $\bt E\ox$. Like the latter, they will be formulated in complex form as
\begin{align}\lab{F}
\bt F\xt=\3E\xt+i\3B\xt.
\end{align}
We begin by interpreting the pulsed beam $\2\Y\xt$ of Section \ref{S:scalar} as a scalar potential and computing a complementary vector potential $\bt A\xt$. Both $\2\Y$ and $\bt A$ are \sl complex. \rm The real field is then obtained as follows. First define the complex fields by complexifying the usual expressions,
\begin{align}\lab{F00}
\bt E=-\grad\2\Y-\pl_t\bt A && \bt B=\curl\bt A,
\end{align}
and let
\begin{align}\lab{F1}
\bt F=\bt E+i\bt B.
\end{align}
Then \eq{F} implies
\begin{align*}
\3E=\re\bt E-\im\bt B && \3B=\re\bt B+\im\bt E.
\end{align*}
Since $\2\Y$ in \eq{Y} is \sl retarded, \rm we require that the four-vector $(\bt A,\2\Y)$ satisfy the \sl Lorenz gauge condition\rm\footnote{Other gauge conditions do not respect causality. While this is acceptable for \sl potentials \rm due to the gauge invariance of the fields, it is not desirable in our case. For example, the fact that $\2\Y$ depends on the complex retarded time $\t-\z$ gave us both the wavefronts $\5O_\x$ and the diffraction semi-hyperboloids $\5H_\h$.
}
\begin{align}\lab{LG}
\div\bt A+\pl_t\2\Y=0.
\end{align}
This relates the unknown vector potential to the known scalar potential. 
We furthermore require that the current density must vanish outside the branch disk $\5D$ of $\z$, which is also the support of the charge density $\2\vr$ defined in \eq{rho}: 
\begin{align}\lab{J0}
4\p\bt J\xt\=\Box\bt A\xt=\30\qq\forall \3x\notin\5D.
\end{align}
Since we expect $\bt A$ to have the same time dependence as $\2\Y$, given by the factor $g(\t-\z)$, we make the \sl Ansatz \rm that
\begin{align}\lab{Av}
\bx{\bt A\xt=\2\Y\xt\3w\ox}
\end{align}
for some static vector field $\3w\ox$.  This will now be used to solve Equations \eq{LG} and \eq{J0}.
Since
\begin{align}\lab{dY}
\grad\2\Y=-\frac{g'(\t-\z)}{\z}\grad\z-\frac{g(\t-\z)}{\z^2}\grad\z\=-\lp\frac{g'}{\z}+\frac{g}{\z^2}\rp\grad\z,
\end{align}
we have
\begin{align*}
\div\bt A=\grad\2\Y\cdot\3w+\2\Y\div\3w=-\lp\frac{g'}{\z}+\frac{g}{\z^2}\rp\grad\z\cdot\3w+\frac{g}{\z}\div\3w.
\end{align*}
Hence \eq{LG} requires
\begin{align*}
-\lp\frac{g'}{\z}+\frac{g}{\z^2}\rp\grad\z\cdot\3w+\frac{g}{\z}\div\3w+\frac{g'}{\z}=0.
\end{align*}
Because this must hold for \sl all \rm analytic signals $g$, we have
\begin{align}\lab{va}
\bx{\div\bt A+\pl_t\2\Y=0\ \forall g\iff \grad\z\cdot\3w=1\ \ \hbox{and}\ \ \div\3w=\frac1{\z}.}
\end{align}
Using \eq{Av} in \eq{J0} gives
\begin{align*}
\Box(\2\Y\3w)=(\Box\2\Y)\3w-2(\grad\2\Y\cdot\grad)\3w-\2\Y\D\3w=\30\ \forall \3x\notin\5D.
\end{align*}
Since $\Box\2\Y=0$ outside $\5D$, the first term vanishes and \eq{dY} gives
\begin{align*}
2\lp\frac{g'}{\z}+\frac{g}{\z^2}\rp(\grad\z\cdot\grad)\3w-\frac{g}{\z}\D\3w=\30\ \forall \3x\notin\5D.
\end{align*}
Again, this must hold for all $g$, hence
\begin{align}\lab{vb}
\bx{(\grad\z\cdot\grad)\3w=\30\ \ \hbox{and}\ \ \D\3w=\30\ \forall \3x\notin\5D.}
\end{align}
Finding $\bt A$ thus reduces to finding a vector field $\3w\ox$ satisfying \eq{va} and \eq{vb}.
Amazingly, although this system appears to be overdetermined, we shall find a reasonable set of solutions $\3w$.

\bf Note: \rm If $\3w=\3w\xt$, then \eq{va} remains the same but \eq{vb} generalizes to
\begin{align}\lab{vbt}
\bx{\pl_t\3w+(\grad\z\cdot\grad)\3w=\30\ \ \hbox{and}\ \ \Box\3w+\frac2\z\pl_t\3w=\30.}
\end{align}

\section{The complex spheroidal frame defined by $\z$}\label{S:frame}

Our main tool for computing $\3w\ox$ will be a complex \sl moving frame \rm \ci{F63} in $\cc3$ which conforms entirely to the spheroidal geometry defined by $\z$. Let
\begin{align*}
\bh\z&=\grad\z=\frac{\3x-i\3a}{\z}=\frac{\r\bh\r+z\bh z-ia\bh z}{\z}=\frac{\r\bh\r+\2z\bh z}{\z},\ \ \hbox{where }\ \ \2z=z-ia.
\end{align*}
This is the complexification of the real unit vector $\bh r=\grad r$, and it satisfies
\begin{align*}
\bh\z^2\=\bh\z\cdot\bh\z=1.
\end{align*}
Note that
\begin{align*}
(\grad\x-i\grad\h)^2=1\imp|\grad\x|^2-|\grad\h|^2=1\ \ \hbox{and}\ \ \grad\x\cdot\grad\h=0,
\end{align*}
confirming that the levelsurfaces $\5O_\x$ and $\5H_\h$ are orthogonal.
Define the complexification $\vq$ of the polar angle $\q$ by
\begin{align}\lab{vq}
\sin\vq=\frac\r{\z},\qq \cos\vq=\frac{\2z}{\z},\qq \sin^2\vq+\cos^2\vq=1,
\end{align}
so that 
\begin{align}\lab{u1}
\bx{\bh\z=\bh\r\sin\vq+\bh z\cos\vq.}
\end{align}
Next, define the complexification of the unit vector $\bh\q=\bh\f\times\bh\z$ by
\begin{align}\lab{u2}
\bx{\bh\vq=\bh\f\times\bh\z=\bh\r\cos\vq-\bh z\sin\vq}
\end{align}
whose real and imaginary parts can be shown to be given by
\begin{align*}
\bh\vq&=\frac{\x r\bh\q+a\h\bh\r}{\x^2+\h^2}+i\frac{\h r\bh\q-a\x\bh\r}{\x^2+\h^2}.
\end{align*}
The three vectors
\begin{align}\lab{u123}
\3u_1=\bh\z,\qq\3u_2=\bh\vq,\qq\3u_3=\bh\f
\end{align}
satisfy the \sl bilinear orthonormality conditions\rm\,\footnote{Not the \sl sesquilinear \rm orthonormality condition $\3u_k^*\cdot\3u_l=\d_{kl}$ based on a \sl hermitian \rm inner product like the one used in quantum mechanics.
}
\begin{align}\lab{con}
\3u_k\cdot\3u_l=\d_{kl}
\end{align}
and form a right-handed \sl complex-orthonormal frame \rm  in $\cc3$.
Every vector field $\3w\ox$ has the unique expansion
\begin{align}\lab{w}
\3w=\sum_k(\3w\cdot\3u_k)\3u_k.
\end{align}
\sl The entire basis is determined by the complex distance $\z$, and thus it conforms completely to the associated complex spheroidal geometry. \rm The use of $\bh\f$ in defining $\bh\vq$ does not spoil this because $\z$ is independent of $\f$. Consequently, the electromagnetic field \eq{F}, \sl including its polarization, \rm will conform to this geometry. As we shall see, that makes it possible to construct \sl coherent \rm fields with vanishing inertia because the polarizations on nearby rays do not clash.

Let the unknown vector field $\3w$ have the expansion  
\begin{align}\lab{vx}
\3w\ox=\a\ox\bh\z+\b\ox\bh\vq+\g\ox\bh\f. 
\end{align}
We want to use Equations \eq{va} and \eq{vb} to compute $(\a,\b,\g)$.
The most difficult one of these would appear to be the first equation in \eq{vb}, \ie
\begin{align}\lab{vb1}
D_\z\3w=\30\ \ \hbox{where}\ \ D_\z=\bh\z\cdot\grad,
\end{align}
since the directional derivative $D_\z$ generally acts on the basis vectors as well as their coefficients. 
Note that the gradient operator can be expressed by
\begin{align}\lab{grad}
\grad=\grad\z\,\pl_\z+\grad\z^*\pl_\z^*+\grad\f\,\pl_\f=\bh\z\pl_\z+\bh\z^*\pl_\z^*+\frac{\bh\f}\r\,\pl_\f
\end{align}
where
\begin{align}\lab{dd*}
\pl_\z=\frac12(\pl_\x+i\pl_\h)&& \pl_\z^*=\frac12(\pl_\x-i\pl_\h)
\end{align}
are the complex derivatives. 
Hence \eq{vb1} becomes
\begin{align}\lab{Dz}
\bx{D_\z\3w=\30 \ \ \hbox{where}\ \ D_\z\=\bh\z\cdot\grad=\pl_\z+|\bh\z|^2\pl_\z^*.}
\end{align}
A glimmer of hope appears when we consider the \sl real \rm limit:
\begin{align*}
\3a\to\30\imp D_r\to \bh r\cdot\grad=\pl_r.
\end{align*}
Since the spherical basis vectors do not depend on $r$, we have
\begin{align}\lab{dr}
\pl_r\bh r=\pl_r\bh\q=\pl_r\bh\f=\30,
\end{align}
and $\pl_r$ does indeed differentiate only the coefficients of a vector field expressed in that basis. 

The complex case is more subtle. While the real variables $r$ and $\q$ are independent, $\z$ and $\vq$ cannot be independent as \sl complex \rm variables since they are both functions of $(\x,\h)$.\footnote{The precise relation between $\z$ and $\vq$ is
\begin{align*}
4a\z\sin\vq&=4a\r=\sr{4a^2+4\x^2}\sr{4a^2-4\h^2}
=\sr{4a^2+(\z+\z^*)^2}\sr{4a^2+(\z-\z^*)^2}.
\end{align*}
}
But somewhat miraculously, it turns out that the essential property \eq{dr} survives complexification.

{\thm {\rm(a)}\label{T:Dz}
The complexified polar angle $\vq$ is constant with respect to differentiation along the complex direction $\bh\z$. That is,
\begin{align}\lab{Dzq}
\bx{D_\z\vq=0.}
\end{align}
{\rm(b)}  Consequently, the basis vectors $\{\bh\z, \bh\vq, \bh\f\}$ are are also constant with respect to $D_\z$:
\begin{align}\lab{Dz1}
D_\z\bh\z= D_\z\bh\vq=D_\z\bh\f=\30.
\end{align}
}\label{T:vq}

\bf Proof: \rm (a) Differentiating $\z\sin\vq=\r$ with respect to $\z^*$ and $\z$ gives
\begin{align}\lab{dzq}
\z\pl_\z^*\vq\cos\vq=\pl_\z^*\r && \sin\vq+\z\pl_\z\vq\cos\vq=\pl_\z\r
\end{align}
From \eq{rz} and \eq{dd*} it follows that
\begin{align*}
\pl_\z^*\r=-\frac{\z^*\2z}{2ia\r}=-\frac{\z^*\z}{2ia\r}\cos\vq &&\pl_\z\r=\frac{\z\2z^*}{2ia\r}.
\end{align*}
The first equation in \eq{dzq} gives
\begin{align}\lab{dzq1}
\pl_\z^*\vq=-\frac{\z^*}{2ia\r}
\end{align}
and the second equation gives
\begin{align*}
\z\pl_{\z}\vq\cos\vq=\frac{\z\2z^*}{2ia\r}-\frac\r{\z}=\frac{\z^2\2z^*-2ia\r^2}{2ia\r\z}.
\end{align*}
Inserting $\z^2=\r^2+\2z^2$ and $-2ia=\2z^*-\2z$, this reduces to
\begin{align*}
\z\pl_\z\vq\cos\vq=\frac{|\2z|^2+\r^2}{\z^*\z}\cdot\frac{\z^*\2z}{2ia\r}
=|\bh\z|^2\frac{\z^*\2z}{2ia\r}=|\bh\z|^2\frac{\z^*\z}{2ia\r}\cos\vq,
\end{align*}
which proves that
\begin{align*}
\pl_\z\vq=|\bh\z|^2\frac{\z^*}{2ia\r}=-|\bh\z|^2\pl_\z^*\vq,\ \ \hbox{hence}\ \ D_\z\vq=0.
\end{align*}
To prove (b), note that
\begin{align*}
\bh\vq=\pl_\vq\bh\z
\end{align*}
by \eq{u1} and \eq{u2}, hence
\begin{align*}
\pl_\z\bh\z=\bh\vq\pl_\z\vq,\qq \pl_\z^*\bh\z=\bh\vq\pl_\z^*\vq
\end{align*}
and
\begin{align}\lab{Dzz}
D_\z\bh\z
=\pl_\z\bh\z+|\bh\z|^2\pl_\z^*\bh\z=\bh\vq\lb\pl_\z\vq+|\bh\z|^2\pl_\z^*\vq\rb=\bh\vq D_\z\vq=\30.
\end{align}
Furthermore, since $\bh\f$ is independent of $\z$ and $\z^*$, we have
\begin{align*}
D_\z\bh\f=\30\ \ \hbox{and}\ \ 
D_\z\bh\vq=D_\z(\bh\f\times\bh\z)=\bh\f\times D_\z\bh\z=\30.\qq\qed
\end{align*}

\section{Computing $\3w\ox$}\label{S:v}

We are now ready to compute
\begin{align}\lab{vx1}
\3w\ox=\a\ox\bh\z+\b\ox\bh\vq+\g\ox\bh\f
\end{align}
by enforcing Equations \eq{va} and \eq{vb}, which are here summarized as
\begin{align}\lab{vab}
\bx{\hbox{(a)}\ \bh\z\cdot\3w=1,\qq\hbox{(b)}\ \div\3w=\frac1{\z},\qq
\hbox{(c)}\ D_{\z}\3w=\30,\qq \hbox{(d)}\ \D\3w=\30.}
\end{align}
\bul Equation \eq{vab} (a) immediately gives $\a\ox\=1$. 

\bul By \eq{Dz}, $D_\z\z=1,\, D_\z\z^*=|\bh\z|^2$, and
\begin{align*}
D_\z\vq^*=\pl_\z\vq^*+|\bh\z|^2\pl_\z^*\vq^*=\frac\z{2ia\r}-|\bh\z|^4\frac\z{2ia\r}\ne0.
\end{align*}
Of the four variables $\z,\z^*,\vq, \vq^*$, only $\vq$ is constant under $D_\z$.
Hence Theorem \ref{T:Dz} shows that $\b$ and $\g$ are functions of the angles $(\vq,\f)$. That is, they are \sl analytic \rm in $\vq$, just as $\2\Y$ is analytic in $\z$:
\begin{align*}
\3w=\bh\z+\b(\vq,\f)\bh\vq+\g(\vq,\f)\bh\f.
\end{align*}
\bul Assume that $\bt A$, like $\2\Y$,  is \sl axisymmetric, \rm \ie its components are independent of $\f$. Then 
\begin{align}\lab{vv0}
\3w=\bh\z+\b(\vq)\bh\vq+\g(\vq)\bh\f
\end{align}
and 
\begin{align*}
D_\z\3w=(D_\z\b)\bh\vq+(D_\z\g)\bh\f=\b' (D_\z\vq)\bh\vq+\g' (D_\z\vq)\bh\f=0
\end{align*}
as required by \eq{vab} (c). We now list some relations that will be needed in solving the remaining equations \eq{vab} (b) and (d).
{\prop
\begin{align*}
&\grad\z=\bh\z&&\curl\bh\z=\30&&\div\bh\z=\frac2{\z}&&\D\z=\frac2{\z}&& \D\bh\z=-\frac{2\bh\z}{\z^2}\\
&\grad\vq=\frac{\bh\vq}{\z}&&\curl\bh\vq=\frac{\bh\f}{\z}&&\div\bh\vq=\frac{\cos\vq}\r&&
\D\vq=\frac{\cot\vq}{\z^2}&&\D\bh\vq=-\frac{\bh\vq+\sin 2\vq\,\bh\z}{\r^2}
\\
&\grad\f=\frac{\bh\f}\r &&\curl\bh\f=\frac{\bh z}\r&&\div\bh\f=0 && \D\f=0&& \D\bh\f=-\frac{\bh\f}{\r^2}.
\end{align*}
}
Proofs of the less obvious results are as follows:
\begin{align*}
\bullet\qq&
\z\cos\vq\grad\vq=\bh\r-(\bh\r\sin\vq+\bh z\cos\vq)\sin\vq=\bh\r\cos^2\vq-\bh\z\cos\vq\sin\vq=\bh\vq\cos\vq\\
\bullet\qq&
\curl\bh\f=\curl(\r\grad\f)=\bh\r\times\grad\f=\frac{\bh\r\times\bh\f}\r=\frac{\bh z}\r\\
\bullet\qq&
\div\bh\vq=\div(\bh\f\times\bh\z)=\bh\z\cdot\curl\bh\f=\frac{\cos\vq}\r\\
\bullet\qq&
\div\bh\f=\div(\r\grad\f)=\bh\r\cdot\bh\f+\r\D\f=0\\
\bullet\qq&
\D\bh\z=\D\grad\z=\grad\D\z=\grad\frac2{\z}=-\frac{2\bh\z}{\z^2}.
\end{align*}
Thus
\begin{align*}
\div\3w=\frac2{\z}+\b\div\bh\vq+\b'\grad\vq\cdot\bh\vq+\g\div\bh\f+\g'\grad\vq\cdot\bh\f
=\frac2{\z}+\frac{\b\cos\vq}\r+\frac{\b'}{\z}
\end{align*}
and Equation \eq{vab} (b) becomes
\begin{align*}
\frac{\b\cos\vq}\r+\frac{\b'}{\z}=-\frac1{\z}.
\end{align*}
To solve this, multiply by $\z\sin\vq=\r$:
\begin{align*}
\b\cos\vq+\b'\sin\vq=\pl_\vq(\b\sin\vq)=-\sin\vq,
\end{align*}
hence
\begin{align}\lab{b0}
\b\sin\vq=\cos\vq+\k
\end{align}
where $\k$ is constant of integration. Thus we have used Equations \eq{vab} (a)--(c) to find
\begin{align}\lab{vv1}
\3w=\bh\z+\b\bh\vq+\g\bh\f\ \ \hbox{where}\ \  \b=\cot\vq+\k\csc\vq.
\end{align}
It remains only to enforce equation \eq{vab} (d). We have
\begin{align*}
\D\3w&=\D\bh\z+(\D\b)\bh\vq+2(\grad\b\cdot\grad)\bh\vq+\b\D\bh\vq+(\D\g)\bh\f+2(\grad\g\cdot\grad)\bh\f+\g\D\bh\f\\
&=\D\bh\z+(\D\b)\bh\vq+2\b'(\grad\vq\cdot\grad)\bh\vq+\b\D\bh\vq+(\D\g)\bh\f+2\g'(\grad\vq\cdot\grad)\bh\f+\g\D\bh\f.
\end{align*}
We find
\begin{align*}
&\b'=-\csc^2\vq-\k\cot\vq\csc\vq,\qqq  \D\b=\frac\b{\r^2},\\
&(\grad\vq\cdot\grad)\bh\vq
=\frac{\bh\vq}{\z}\cdot\lp\bh\z\pl_\z+\bh\z^*\pl_\z^*+\frac{\bh\f}\r\,\pl_\f\rp
=\frac{\bh\vq\cdot\bh\z^*}{\z}\,\pl_{\z}^*\bh\vq=-\frac{2ia\r}{\z^*\z^2}\,\pl_{\z}^*\bh\vq.
\end{align*}
But
\begin{align*}
\pl_{\z}^*\bh\vq=(\pl_{\z}^*\vq)(\pl_\vq\bh\vq)=\frac{\z^*}{2ia\r}\bh\z,
\end{align*}
hence
\begin{align*}
(\grad\vq\cdot\grad)\bh\vq=-\frac{\bh\z}{\z^2}.
\end{align*}
Since $(\grad\vq\cdot\grad)\bh\f=\30$, we have
\begin{align*}
\D\3w&=-\frac2{\z^2}\bh\z+\frac\b{\r^2}\bh\vq-\frac{2\b'}{\z^2}\bh\z
-\frac\b{\r^2}\lb\bh\vq+\bh\z\sin 2\vq\rb +\lb\D\g-\frac\g{\r^2}\rb\bh\f.
\end{align*}
By yet another miracle, the coefficients of $\bh\z$ and $\bh\vq$ both vanish and we are left with
\begin{align}\lab{Dv1}
\D\3w=\lb\D\g-\frac\g{\r^2}\rb\bh\f.
\end{align}
Thus \eq{vab} (d) requires that $\g$ satisfy the same equation as $\b$, for which we have already found a general solution:
\begin{align}\lab{Dv2}
\bx{\D\3w=\30\imp\D\g=\frac\g{\r^2}\imp\g=\l\cot\vq+\m\csc\vq.}
\end{align}
We have proved the following result.

{\thm The general axisymmetric solution of Equations {\rm\eq{vab}} is
\begin{align}\lab{soln}
\3w=\bh\z+(\cot\vq+\k\csc\vq)\bh\vq+(\l\cot\vq+\m\csc\vq)\bh\f
\end{align}
where $\k,\l,\m$ are arbitrary complex constants.
}\label{T:w}

\bf Remark. \rm The solution \eq{soln} is singular on the branch disk $\5D$ since $\z$, hence also $\vq$, is discontinuous there. In fact, 
\begin{align*}
\x\to0,\ z\to 0^\pm&\imp\z\to\mp i\sr{a^2-\r^2},\qq \2z\to-ia\\
&\imp\cos\vq\to\pm \frac{a}{\sr{a^2-\r^2}},\qq\sin\vq\to\pm i\frac\r{\sr{a^2-\r^2}}
\end{align*}
In addition, $\3w$ and thus $\bt A$, is singular on the $z$-axis. To bring out this singularity, recall that $\sin\vq=\r/\z$, hence \eq{soln} can be rewritten as
\begin{align}\lab{vv}
\bx{\3w=\bh\z+\frac{\z}\r(\cos\vq+\k)\bh\vq+\frac{\z}\r(\l\cos\vq+\m)\bh\f.}
\end{align}

\section{The general pulsed-beam fields $\bt F_\pm$}\label{gensol}

According to \eq{Av} and \eq{vv}, the general axisymmetric vector potential complementing $\2\Y$ is
\begin{align}\lab{A0}
\bt A=\frac{g}{\z}\bh\z+\frac{g}\r(\cos\vq+\k)\bh\vq+\frac{g}\r(\l\cos\vq+\m)\bh\f,
\end{align}
where we have used the shorthand $g\=g(\t-\z)$ and $\k,\l,\m$ are arbitrary complex constants. 

Let us compute the complex electric and magnetic fields defined in \eq{F1}:
\begin{align*}
\bt E&=-\grad\2\Y-\pl_t\bt A\\
&=\lb\frac{g'}{\z}+\frac{g}{\z^2}\rb\bh\z-\frac{g'}{\z}\bh\z
-\frac{g'}\r(\cos\vq+\k)\bh\vq-\frac{g'}\r(\l\cos\vq+\m)\bh\f.
\end{align*}
Note that the radiating longitudinal component $(g'/\z)\bh\z$ cancels, leaving only the non-radiating longitudinal component $(g/\z^2)\bh\z$:
\begin{align}\lab{E2}
\bx{\bt E=\frac{g}{\z^2}\bh\z-\frac{g'}\r(\cos\vq+\k)\bh\vq-\frac{g'}\r(\l\cos\vq+\m)\bh\f.}
\end{align}
The radiation is seen to be confined to the (complex) transversal directions $\bh\vq$ and $\bh\f$. The $1/\r$ decay of the radiating terms indicates that the singularity along the $z$-axis carries a charge-current distribution and thus represents a \sl filament \rm attached to the branch disk. This will be investigated in detail elsewhere.

For computing $\bt B=\curl\bt A$, it will be convenient to rewrite $\bt A$ as
\begin{align}\lab{A1}
\bt A=\frac{g}{\z}\bh\z+g(\cot\vq+\k\csc\vq)\grad\vq+g(\l\cos\vq+\m)\grad\f,
\end{align}
where we have used $\bh\vq=\z\grad\vq,\, \bh\f=\r\grad\f$, and $\z/\r=\csc\vq$. Then
\begin{align*}
\bt B&=-g'(\cot\vq+\k\csc\vq)\bh\z\times\grad\vq-g'(\l\cos\vq+\m)\bh\z\times\grad\f-\l g\sin\vq\grad\vq\times\grad\f\\
&=-\frac{g'}{\z}(\cot\vq+\k\csc\vq)\bh\f+\frac{g'}\r(\l\cos\vq+\m)\bh\vq-\frac{\l g\sin\vq}{\r\z}\bh\z,
\end{align*}
whose radiative structure is seen more clearly in the form
\begin{align}\lab{B2}
\bx{\bt B=-\frac{\l g}{\z^2}\bh\z+\frac{g'}\r(\l\cos\vq+\m)\bh\vq-\frac{g'}\r(\cos\vq+\k)\bh\f.}
\end{align}
Note the close similarities between $\bt E$ and $\bt B$, down to the non-radiating longitudinal parts. These similarities combine to give 
\begin{gather}\lab{FF}
\bx{\bt F_\pm\=\bt E\pm i\bt B
=p_\pm\frac{g}{\z^2}\bh\z+(q_\pm-p_\pm\cos\vq)\frac{g'}\r\bt\vf_\pm}
\end{gather}
where
\begin{gather}\lab{phitil}
\bx{p_\pm=1\mp i\l, \qq q_\pm=-\k\pm i\m,\qq\bt\vf_\pm=\bh\vq\pm i\bh\f.}
\end{gather}
Note that $\bt\vf_+$ and $\bt\vf_-$ are \sl null vectors: \rm
\begin{align}\lab{nullity}
\bt\vf_\pm^2=\bh\vq^2-\bh\f^2\pm 2i\bh\vq\cdot\bh\f=0.
\end{align}
They form a \sl complex basis of helicity $\pm$ \rm for the the directions $\bh\vq, \bh\f$ orthogonal to $\bh\z$, and will play an important role in the representation of our coherent wavelets. Note also that
\begin{align}\lab{ypm0}
\frac{\bt\vf_\pm}\r=\frac{\grad\vq}{\sin\vq}\pm i\grad\f=\grad\y_\pm
\end{align}
where
\begin{align}\lab{ypm}
\y_\pm(\vq,\f)=\ln\tan(\vq/2)\pm i\f=\ln\lp e^{\pm i\f}\tan(\vq/2)\rp,
\end{align}
hence 
\begin{align}\lab{dvf}
\bx{\r\inv\bt\vf_\pm\cdot\dd\3x=\frac{\dd\vq}{\sin\vq}\pm i\dd\f=\dd\y_\pm.}
\end{align}
This can be used to write
\begin{align}\lab{Fdx}
\bx{\bt F_\pm\cdot\dd\3x
=p_\pm\frac{g}{\z^2}\dd\z+(q_\pm-p_\pm\cos\vq)g'\dd\y_\pm.}
\end{align}

\bf Remark 1. \rm
It might seem that using both complex fields $\bt F_\pm$ in \eq{FF} is superfluous. But since $\bt E$ and $\bt B$ are \sl complex, \rm $\bt F_+$ and $\bt F_-$ are not related by complex conjugation, thus  both fields are needed to recover $\bt E$ and $\bt B$:
\begin{align*}
2\bt E=\bt F_++\bt F_-&&2i\bt B=\bt F_+-\bt F_-.
\end{align*}
On the other hand, to compute \sl real \rm fields, we need \sl either \rm $\bt F_+$ or $\bt F_-$ but not both. Define the two real fields $(\3E_+,\3B_+)$ and $(\3E_-,\3B_-)$ by
\begin{align*}
\bt F_\pm=\3E_\pm\pm i\3B_\pm,
\end{align*}
so that
\begin{align}\lab{EB}
\3E_\pm&=\re\bt F_\pm=\re\bt E\mp\im\bt B\\
\3B_\pm&=\pm\im\bt F_\pm=\re\bt B\pm\im\bt E\,. \nt
\end{align}
Clearly these two fields are inequivalent, so $\bt F_+$ and $\bt F_-$ give two independent real solutions to Maxwell's equations. Each of these solutions corresponds to a \sl class \rm of equivalent complex fields. Namely, the real field $(\3E_\pm, \3B_\pm)$ remains unchanged when 
\begin{align}\lab{Ep}
\bt E\to\bt E+\bt E'_\pm\ \ \hbox{and}\ \ \bt B\to\bt B+\bt B'_\pm\ \ \hbox{where}\ \ \bt B'_\pm=\pm i\bt E',
\end{align}
since this leaves $\bt F_\pm$ invariant:
\begin{align*}
\bt F_\pm\to(\bt E+\bt E'_\pm)\pm i(\bt B\pm i\bt E'_\pm)=\bt E\pm i\bt B=\bt F_\pm.
\end{align*}
If the changes are induced by changes $(\bt A'_\pm, \2\Y'_\pm)$ of the potentials, then 
\begin{align}\lab{AY'}
-\grad\2\Y'_\pm-\pl_t\bt A'_\pm=\bt E'_\pm=\mp i\bt B'_\pm=\mp i\curl\bt A'_\pm.
\end{align}
Taking the divergence of both sides and using the Lorenz condition gives
\begin{align}\lab{cxgauge}
-\D\2\Y'_\pm-\pl_t\div\bt A'_\pm=-\D\2\Y'_\pm+\pl_t^2\2\Y'_\pm=\Box\2\Y'_\pm=0.
\end{align}
Similarly, taking the curl of \eq{AY'} yields $\Box\bt A_\pm=\30$. The addition of a sourceless term $(\bt A_\pm',\2\Y'_\pm)$ to the 4-potential $(\bt A,\2\Y)$ thus leaves the real field $(\3E_\pm, \3B_\pm)$ invariant. However, this \sl does \rm change the field $(\3E_\mp, \3B_\mp)$ of opposite helicity:
\begin{align*}
&\bt F_\mp\to(\bt E+\bt E'_\pm)\mp i(\bt B\pm i\bt E'_\pm)=\bt F_\mp+2\bt E'_\pm\\
\imp&\3E_\mp\to\3E_\mp+2\re\bt E',\qq \3B_\mp\to\3B_\mp\mp2\im\bt E'.
\end{align*}
Furthermore, note that $(\3E_\pm, \3B_\pm)$ remains unaffected outside its source region even if $(\bt A_\pm',\2\Y'_\pm)$ has a charge-current density whose support is contained in that of $(\bt A, \2\Y)$. This is a new kind of `gauge freedom' belonging to the complex representation \eq{FF}.
In Secton \ref{S:KN} we shall encounter an example of a `pure gauge' field where $\bt A\ne\30$ but $\bt F_\pm=\30$; see \eq{nullnull}.

\bf Remark 2. \rm Equation \eq{FF} shows that the \sl radiating \rm parts of the two independent fields $(\3E_+,\3B_+)$ and $(\3E_-,\3B_-)$ associated with $\bt F_+$ and $\bt F_-$ have helicity $+$ and $-$, respectively. This confirms that both fields $\bt F_\pm$ are needed, as explained above.

\section{Coherent electromagnetic wavelets}\label{S:coherent}

The inertia density \eq{I} of the field $\bt F_\pm$ is
\begin{align*}
\5I_\pm=\frac12|\bt F_\pm^2|.
\end{align*} 
Since $\bh\z\cdot\bt\vf_\pm=0$ and $\bt\vf_\pm^2=0$, \eq{FF} gives
\begin{align}\lab{FF2}
\bt F_\pm^2=p_\pm^2\frac{g(\t-\z)^2}{\z^4}.
\end{align}
Thus \sl one, but not both, of the fields $\bt F_+$ and $\bt F_-$ can be made null: \rm 
\begin{align}\lab{FF3}
\l=-i&\imp p_+=0\imp\bt F_+=q_+\frac{g'(\t-\z)}\r\,\bt\vf_+\imp \bt F_+^2=0\\
\l=i&\imp p_-=0\imp\bt F_-=q_-\frac{g'(\t-\z)}\r\,\bt\vf_-\imp \bt F_-^2=0.\nt
\end{align}
Up to a complex scalar factor, we have thus found two \sl unique \rm null fields.

{\thm The field $\bt F_\pm$ is null if and only if $\l=\mp i$, and then  
\begin{align}\lab{coh}
\bx{\bt F_\pm\xt=\frac{g'(\t-\z)}\r\,\bt\vf_\pm,\qqq\bt F_\pm\cdot\dd\3x=g'(\t-\z)\dd\y_\pm}
\end{align}
or a complex scalar multiple thereof, where $\y_\pm$ is given by {\rm\eq{ypm}}.
}\label{T:coh}

\defin
We call $\bt F_\pm$ the \sl coherent electromagnetic wavelets with helicity $\pm$ and driving signal $g$. \rm

Recall that $\bt A$ contained three independent complex parameters $\k,\l,\m$. To make $\bt F_\pm$ null, we must choose $\l=\mp i$, respectively. That leaves two complex parameters, given now by $q_\pm$ in \eq{FF3}. To reconstruct the \sl complex \rm fields $\bt E$ and $\bt B$, we need both the null field $\bt F_\pm$ and the non-null field $\bt F_\mp$, hence both $q_+$ and $q_-$. However, if we are interested only in \sl null \rm fields, then we need only $q_+$ (for positive helicity) or $q_-$ (for negative helicity).

{\thm $\bt F_\pm$ is an an electromagnetic pulsed beam propagating in the direction of $\3a$, with wavefronts $\5O_\x$ exploding as $\x=ct$, and diffracting along $\5H_\h$. 
\label{T:EMPB}
}

\bf Proof: \rm Apply the argument used with Equation \eq{g} to interpret $\2\Y$ as a scalar pulsed beam to the \sl pulse-shape factor \rm in \eq{coh}:
\begin{align*}
g'(\t-\z)=-\frac i{2\p}\int_0^\infty\dd\o\,\o e^{-i\o(t-\x)}e^{-\o(s-\h)}\1g_0\0\o.\qq\qed
\end{align*}

\bf Remark. \rm Since $\bt\vf_\pm=\bh\vq\pm i\bh\f$ and
\begin{align*}
\bh\vq=\frac{\2z\bh\r-\r\bh z}\z,
\end{align*}
by \eq{u2} and \eq{vq}, $\bt F_\pm$ is singular on the branch disk $\5D$ of $\z$. Furthermore, \eq{coh} shows that it is singular along the $z$-axis $\r=0$. $\bt F_\pm$ is analytic elsewhere, and this implies that its charge-current distribution is supported on $\5D$, representing a spinning charged disk, as well as the $z$-axis, representing \sl vortex jets \rm generated by the spin of $\5D$ in the positive and negative $z$-directions. However, as shown in Theorem \ref{T:EMPB}, the jet along the negative $z$-axis is much weaker due to the pulse-shape factor $g'(\t-\z)$.

The sources of $\bt F_\pm$ will be studied elsewhere.

\section{The real and complex null congruences of $\bt F_\pm$}\label{S:CxCongr}

The null fields \eq{FF3} define both real and complex twisted null congruences in spacetime. The real congruence follow from the nullity condition
\begin{align*}
(\3E_\pm\pm i\3B_\pm)^2=0\imp \3S_\pm^2\=|\3E_\pm\times\3B_\pm|^2=u_\pm^2\=\frac14(\3E_\pm^2+\3B_\pm^2)^2,
\end{align*}
which shows that the energy propagates at the speed of light:
\begin{align*}
\3v_\pm=\frac{\3S_\pm}{u_\pm}\imp
\3v_\pm^2=\frac{\3S_\pm^2}{u_\pm^2}=1\=c^2.
\end{align*}

{\thm The energy flow velocity for the coherent wavelets {\rm\eq{FF3}} is 
\begin{align}\lab{vreal}
\3v_\pm\ox&=\frac\x a\sr{\frac{a^2-\h^2}{\x^2+a^2}}\,\bh\r+\frac\h a\,\bh z\pm \sr{\frac{a^2-\h^2}{\x^2+a^2}}\,\bh\f\\
&=\bh\x\sr{\frac{\x^2+\h^2}{\x^2+a^2}}\pm\bh\f\sr{\frac{a^2-\h^2}{\x^2+a^2}}\,,\nt
\end{align}
which is identical to the velocity field {\rm\eq{upm}--\eq{upmos}} associated with the scalar wavelet $\2\Y$. The 4-vector field $v^\m_\pm=(\3v_\pm, 1)$ thus forms a twisting congruence in Minkowski spacetime. This can be expressed as a differential form:
\begin{align}\lab{vreal1}
\3v_\pm\ox\cdot\dd\3x=\frac{\x^2+\h^2}{\x^2+a^2}\,\dd\x\pm\frac{a^2-\h^2}a\,\dd\f.
\end{align}
}\label{T:vCoh}

\bf Proof: \rm Recall from \eq{u1} and \eq{u2} that
\begin{align*}
\z\bh\z=\r\bh\r+\2z\bh z&&\z\bh\vq=\2z\bh\r-\r\bh z&& \2z=z-ia.
\end{align*}
It follows that
\begin{align*}
&|\z|^2\bh\vq^*\cdot\bh\vq=r^2+a^2=|\z|^2|\bh\z|^2=\x^2-\h^2+2a^2\\
& |\z|^2\bh\vq^*\times\bh\vq=2ia\r\bh\f.
\end{align*}
To simplify the notation, let
\begin{align*}
X=\x^2+a^2\ \ \hbox{and}\ \  Y=a^2-\h^2
\end{align*}
so that
\begin{align*}
XY=a^2\r^2&&X+Y=\x^2-\h^2+2a^2 && X-Y=\x^2+\h^2=|\z|^2.
\end{align*}
From $\bt\vf_\pm=\bh\vq\pm i\bh\f$ it follows that
\begin{align*}
&|\z|^2\bt\vf_\pm^*\cdot\bt\vf_\pm=2X\\
&|\z|^2\bt\vf_\pm^*\times\bt\vf_\pm=\pm 2i\lb\x\r\bh\r+(X\h/a)\bh z\pm a\r\bh\f\rb.
\end{align*}
For the null field $\bt F_\pm=(g'/\r)\bt\vf_\pm$ we have 
\begin{align*}
u_\pm&=\frac12\bt F_\pm^*\cdot\bt F_\pm=\frac{| g'|^2}{\r^2}\,\frac{2X}{X-Y}\\
\3S_\pm&=\pm\frac{\bt F_\pm^*\times\bt F_\pm}{2i}
=\frac{|g'|^2}{\r^2}\,\frac{\x\r\bh\r+(X\h/a)\bh z\pm a\r\bh\f}{X-Y},
\end{align*}
hence 
\begin{align*}
\3v_\pm\ox=\frac{\x\r}X\,\bh\r+\frac\h a\,\bh z\pm \frac {a\r}X\,\bh\f\imp  \3v_\pm^2=1.
\end{align*}
Using $a\r=\sr{XY}$, we obtain \eq{vreal}. To derive \eq{vreal1}, use the identities
\begin{align*}
|\grad\x|=\sr{\frac{\x^2+\h^2}{\x^2+a^2}},\qq
\bh\f=\r\grad\f=\frac{\sr{(\x^2+a^2)(a^2-\h^2)}}a\grad\f.\qq \qed
\end{align*}

Note that $\3v_\pm$ is time-independent because the pulse factors $|g'|^2$ in $\3S_\pm$ and $u_\pm$ cancel. Therefore the integral curves of $\3v_\pm$ are straight lines in $\rr3$ propagating at the speed of light. Accompanied by their fields $(\3E_\pm, \3B_\pm)$, these can now be justifiably interpreted as \sl light rays. \rm

Whereas the physical interpretation of the null congruence associated with the scalar wavelet $\2\Y$ was vague because scalar (longitudinal) waves would seem to propagate in the direction orthogonal to their wavefronts, so their congruence could not be twisted, the identical null congruence defined by $\3v_\pm$ has a clear physical interpretation: \sl the spinning disk $\5D$ emits light rays whose trajectories are given by \rm
\begin{gather*}
\3x_\pm\0\x=\3\r_0+t\3v^o_\pm,\qq \3\r_0\in\5D,\qq t\ge 0\\
\hbox{where}\ \ \3v^o_\pm=\sgn z\frac{\sr{a^2-\r_0^2}}a\,\bh z\pm\frac{\r_0}a\,\bh\f.
\end{gather*} 
The rays with $\r_0=0$ form vertical \sl vortex jets \rm along the $z$-axis, while the rays emanating from any circle of radius $0<\r_0<a$ generate the hyperboloid $\5H_{\h_0}\cup\5H_{-\h_0}$, with opposite rulings on the upper and lower halves. The rays coming from the circle $\r_0=a$ are tangent to that circle, forming the degenerate hyperboloid $\5H_0$.

To find the \sl complex \rm congruence of $\bt F_\pm$, define the complex electromagnetic energy and momentum densities
\begin{align}\lab{CxES}
\2u=\frac12(\bt E^2+\bt B^2),\qq \bt S=\bt E\times\bt B.
\end{align}
Letting
\begin{align*}
L=\cos\vq+\k,\qq M=\l\cos\vq+\m
\end{align*}
in \eq{E2} and \eq{B2}, we obtain
\begin{align}\lab{CxES1}
\2u&=\frac{1+\l^2}2\,\frac{g^2}{\z^4}+\frac{g'^2}{\r^2}(L^2+M^2)\\
\bt S&=\frac{g'^2}{\r^2}(L^2+M^2)\bh\z+\frac{gg'}{\z^2\r}(L+\l M)\bh\vq+
\frac{gg'}{\z^2\r}(M-\l L)\bh\f.\nt
\end{align}
Now assume that $\bt F_\pm$ is null. Then $\l=\mp i$ and
\begin{align*}
L^2+M^2=\k^2+\m^2=q_+ q_-,\  L+\l M=-q_\pm,\  M-\l L=\pm iq_\pm,
\end{align*}
so
\begin{align}\lab{CxES2}
\2u&=q_+ q_-\frac{g'^2}{\r^2},\qq
\bt S=q_+ q_-\frac{g'^2}{\r^2}\bh\z-q_\pm\frac{gg'}{\z^2\r}\,\bt\vf_\pm.
\end{align}
Hence the \sl complex energy propagation velocity \rm is 
\begin{align}\lab{Cxv}
\bx{\bt v_\pm\=\frac{\bt S}{\2u}=\bh\z-q_\mp\inv\frac{g\r}{g'\z^2}\,\bt\vf_\pm\,.}
\end{align}
We shall suppress the helicity index $\pm$ and write $\bt v$ for notational simplicity.
Thus
\begin{align*}
\bt v^2=\bh\z^2=1\=c^2,
\end{align*}
so the corresponding complex energy flow velocity 4-vector
\begin{align*}
\2v^\m=(\bt v, 1),\qq v_\m=(-\bt v, 1)
\end{align*}
defines a null congruence in $\cc4$:
\begin{align}\lab{CxNullCong}
v_\m v^\m=0.
\end{align}
To show that this congruence is \sl twisted, \rm consider the complex 1-form
\begin{align}\lab{vdx}
\2v&\=\2v_\m \dd x^\m=\dd t-\bt v\cdot\dd\3x=\dd t-\dd\z+q_\mp\inv\frac{g\r^2}{g'\z^2}\,\dd\y_\pm,
\end{align}
where we have used \eq{dvf}. Since $\t=t-is$ and $s$ is fixed, 
\begin{align}\lab{ret}
\dd t-\dd\z=\dd\t_\z \ \ \hbox{where}\ \ \t_\z=\t-\z
\end{align}
is the \sl complex retarded time. \rm
Letting
\begin{align*}
h(\t_\z)=q_\mp\inv\frac{g(\t_\z)}{g'(\t_\z)}
\end{align*}
and recalling that $\r/\z=\sin\vq$, we have
\begin{align}\lab{2v}
\bx{\2v=\dd\t_\z+h\sin\vq\,\dd\vq\pm ih\sin^2\vq\,\dd\f.}
\end{align}
The exterior derivative \ci{F63} of $\2v$ is given by
\begin{align*}
\dd\2v&=\sin\vq\,\dd h\wedge\dd\vq\pm i\sin^2\vq\,\dd h\wedge\dd\f\pm ih\sin(2\vq)\,\dd\vq\wedge\dd\f,
\end{align*}
hence
\begin{align*}
\2v\wedge\dd\2v=\pm ih\sin(2\vq)\,\dd\t_\z\wedge\dd\vq\wedge\dd\f.
\end{align*}
Since $g(\t_\z)$ is analytic, its zeros are isolated and so are those of $h(\t_\z)$.
It follows that the congruence is \sl twisting \rm wherever $\sin(2\vq)\ne0$. But
\begin{align*}
\sin(2\vq)=2\sin\vq\cos\vq=\frac{2\r\2z}{\z^2}=0\imp\r=0
\end{align*}
since $\2z=z-ia\ne 0$ for all $\3x$ if $a>0$. The complex congruence is therefore twisting everywhere outside the $z$-axis. Recall from \eq{twist1} that the \sl real \rm congruence defined by $\3E_\pm\pm i\3B_\pm$ was twisting everywhere outside the $xy$-plane. 

\bf Remark. \rm The complex electromagnetic energy-momentum density is related to the real fields $(\3E_\pm, \3B_\pm)$ as follows:
\begin{align*}
\2u&=\frac12(\3E_+\cdot\3E_-+\3B_+\cdot\3B_-)+\frac i2(\3E_-\cdot\3B_+-\3E_+\cdot\3B_-)\\
\bt S&=\frac12(\3E_+\times\3B_-+\3E_- \times\3B_+)+\frac i2(\3E_+\times\3E_-+\3B_+\times\3B_-)
\end{align*}
They represent a Hermitian \sl cross-correlation \rm between the fields  $(\3E_+, \3B_+)$ and $(\3E_-, \3B_-)$ whose physical significance, if any, remains to be investigated.

\section{Relation to Kerr-Newman black holes}\label{S:KN}

The Kerr-Newman solution of the Einstein-Maxwell equations in general relativity can be described in a simple way as follows \ci{Wik-KN}. 

\bul The spacetime metric $g^\pm_{\m\n}$ of the solution is given in terms of flat Minkowski spacetime coordinates $(x,y,z,t)$ by
\begin{align}\lab{KN}
g^\pm_{\m\n}=\h_{\m\n}+f k^\pm_\m k^\pm_\n
\end{align}
where $\h_{\m\n}={\rm diag}(1,-1,-1,-1)$ is the Minkowski metric, the scalar function $f$ is given by
\begin{align}\lab{KNf}
f=G\frac{r^2(2Mr-Q^2)}{r^4+a^2z^2}
\end{align}
where $G$ is Newton's gravitational constant, and
\begin{align}\lab{KNk}
k^\pm_\m=\lp\3k_\pm,1\rp\ \ \hbox{where}\ \ \3k_\pm=\lp\frac{rx\mp ay}{r^2+a^2},\frac{ry\pm ax}{r^2+a^2},\frac zr\rp.
\end{align}
This describes a spinning black hole with mass $M$, charge $Q$, and angular momentum $\3L=mca\bh z$. It can be checked that $\3k_\pm^2=1$, hence $k_\m k^\m=0$ and $k_\m$ defines a null congruence. Since $\3k_\pm$ is time-independent, the rays of this congruence are straight lines in Minkowski space. They become \sl geodesics \rm in the curved space defined by $g^\pm_{\m\n}$.

\bul The electromagnetic field $F_{\m\n}$ of the Kerr-Newman black hole is derived from the 4-vector potential
\begin{align}\lab{KNA}
A_\m=\frac{Qr^3}{r^4+a^2z^2}\, k^\pm_\m,\qq F_{\m\n}=\pl_\n A_\m-\pl_\m A_\n,
\end{align}
where we have suppressed the helicity index $\pm$ in $A_\m$ and $F_{\m\n}$.
The parameter $a$ above coincides with our parameter $a$, which represents the magnitude of the imaginary displacement of the point source. However, the parameter $r$ does \sl not \rm signify the Euclidean distance $|\3x|$ in $\rr3$. Instead, it identical to our parameter $\x$, representing the real part of the complex distance $\z=\x-i\h$. Since $az=\x\h$ by \eq{rz}, $f$ and $A_\m$ simplify to
\begin{align}\lab{fA}
f=G\frac{2M\x-Q^2}{\x^2+a^2},\qq A_\m=\frac{Q\x}{\x^2+a^2}\, k^\pm_\m.
\end{align}

{\thm The 4-vector $k^\pm_\m$ defining the null congruence of the Kerr-Newman solution {\rm \eq{KN}} is identical with the 4-velocities {\rm\eq{upm}} and {\rm\eq{vreal}} defining the null congruences of the scalar wavelet $\2\Y$ {\rm\eq{Y}} and the coherent electromagnetic wavelets $\bt F_\pm$ {\rm\eq{coh}}, respectively.
}

\bf Proof: \rm Since
\begin{align*}
\mp ay\bh x\pm ax\bh y=\pm a\r\bh\f \ \ \hbox{and}\ \ \frac z\x=\frac\h a,
\end{align*}
\eq{KNk} gives
\begin{align*}
\3k_\pm=\frac{\x\3\r\pm a\r\bh\f}{\x^2+a^2}+\frac\h a\,\bh z&=\frac{\x\r}{\x^2+a^2}\,\bh\r\pm\frac{a\r}{\x^2+a^2}\,\bh\f+\frac\h a\,\bh z.
\end{align*}
By  \eq{rz} we have $a\r=\sr{\x^2+a^2}\sr{a^2-\h^2}$, hence
\begin{align*}
\3k_\pm=\frac\x a\sr{\frac{a^2-\h^2}{\x^2+a^2}}\,\bh\r\pm\sr{\frac{a^2-\h^2}{\x^2+a^2}}\,\bh\f+\frac\h a\,\bh z.\qq\qed
\end{align*}

However, note that the electromagnetic field $\bt F_\pm$ of our coherent wavelets  does \sl not \rm coincide with that of the Kerr-Newman solution.\footnote{The complex representation of $F_{\m\n}$ corresponding to $\bt F_\pm$ is given in 4D spacetime notation by the field
\begin{align*}
\2F^\pm_{\m\n}=F_{\m\n}\pm i*F_{\m\n},
\end{align*}
where $*F_{\m\n}=\frac12\e_{\m\n\a\b}F^{\a\b}$ is the (Hodge) dual of $F_{\m\n}$. Since $**=-1$, $\2F_{\m\n}$ is \sl self-dual: \rm 
\begin{align*}
\pm i*\2F^\pm_{\m\n}=\2F^\pm_{\m\n}.
\end{align*}
}
In particular, $\bt F_\pm$ is \sl pulsed \rm while $F_{\m\n}$ in \eq{KNA} is \sl stationary. \rm Both fields conform to the oblate spheroidal geometry of the Kerr-Newman spacetime. It remains to be seen how our coherent wavelets fit in with the theory of spinning black holes. 

\sl Could a pulsed, radiating version $\2g_{\m\n}$ of the Kerr-Newman metric \eq{KN} exist which complements $\bt F_\pm$? \rm If so, this would give the following sequence of extensions:
\begin{align*}
\hbox{Scalar wavelet $\2\Y\to$ EM wavelet $\bt F_\pm\to$ gravitational wavelet $\2g^\pm_{\m\n}$\,.} 
\end{align*}

\bf Remark. \rm Recall the general form of the axisymmetric complex vector potential $\bt A$ complementing $\2\Y$:
\begin{align*}
\bt A\xt=\2\Y\xt\3w\ox,\qq  \3w=\bh\z+\frac{\z}\r(\cos\vq+\k)\bh\vq+\frac{\z}\r(\l\cos\vq+\m)\bh\f.
\end{align*}
The complex 4-potential $(\bt A, \2\Y)$ for the general EM wavelet $\bt F_\pm=\bt E\pm i\bt B$ (not necessarily null) can thus be written in the form \eq{fA} as
\begin{align*}
\2A_\m\xt=\2\Y\xt w_\m\ox\ \ \hbox{where}\ \ w_\m=(\3w, 1).
\end{align*}
$w_\m$ defines a complex null congruence if and only if $\3w^2\=1$, or
\begin{align*}
&(\cos\vq+\k)^2+(\l\cos\vq+\m)^2\=0\iff 1+\l^2=\k+\l\m=\k^2+\m^2=0.
\end{align*}
The conditions for $w_\m$ to define a complex null congruence are therefore
\begin{align}\lab{nullnull}
\l=\mp i\ \ \hbox{and}\ \   \k=\pm i\m,\ \ \hbox{hence}\ \ \3w=\bh\z+\frac\z\r(\cos\vq+\k)\bt\vf_\mp.
\end{align}
By \eq{FF}, this implies $p_\pm=q_\pm=0$, hence $\bt F_\pm\=\30$.

That is, $w_\m$ \sl defines a complex null congruence if and only if the field $\bt F_\pm$ vanish identically. \rm This can be checked directly by computing
\begin{align*}
\bt E=\frac g{\z^2}\bh\z-\frac{g'}\r(\cos\vq+\k)\bt\vf_\mp\ \ \hbox{and}\ \  \bt B=\mp i\bt E.
\end{align*}
This is an example of a `pure gauge' field in the complex sense of \eq{Ep}, where $\bt A\ne\30$ but $\bt F_\pm=\30$.

Nevertheless, the null fields $\bt F_\pm$ defined by $\2A_\m$ with $\l=\mp i$ and $\k\ne\pm i\m$ define real and complex twisting null congruences (Section \ref{S:CxCongr}), and this the important fact since potentials have no direct physical significance.

\section*{Acknowledgements}
I thank Iwo Bialynicki-Birula, Richard Matzner, Ted Newman, Ivor Robinson and Andrzej Trautman for helpful discussions of null fields and twisted congruences, and Arthur Yaghjian for an informative discussion of electromagnetic propagation speeds. This work was supported by AFOSR Grant \#FA9550-08-1-0144.

\end{document}